\documentclass{article}

\usepackage{arxiv}

\usepackage[utf8]{inputenc} 
\usepackage[T1]{fontenc}    
\usepackage{hyperref}       
\usepackage{url}            
\usepackage{booktabs}       
\usepackage{amsmath}
\usepackage{amssymb}
\usepackage{color}
\usepackage{amsfonts}
\usepackage{nicefrac}       
\usepackage{microtype}      
\usepackage{cleveref}       
\usepackage{lipsum}         
\usepackage{graphicx}
\usepackage{doi}

\usepackage{algorithm}          
\usepackage{algcompatible}

\usepackage{psfrag}
\usepackage{indentfirst}
\usepackage{cite}
\usepackage{subfigure}
\usepackage{balance}
\usepackage{placeins}
\usepackage{multirow}
\usepackage{booktabs}
\usepackage{epstopdf}
\usepackage{arydshln}

\newtheorem{theorem}{Theorem}

\newtheorem{definition}{Definition}
\newtheorem{assumption}{Assumption}
\newtheorem{remark}{Remark}

\def\Re{{\mathop{\Bbb R}}}
\def\Na{{\mathop{\Bbb N}}}

\def\dsst{\displaystyle}

\def\qed{\hfill \square}

\def\ARGMIN{\mathop{{\rm argmin}}}

\def\eqref#1{(\ref{#1})}

\def\qed{\hfill\ensuremath{\Box}}

\newcommand{\cK}{{\mathcal K}}
\newcommand{\cKL}{{\mathcal KL}}

\newcommand\Oplus{\scalebox{1.5}{$\oplus$}}

\title{Robust Moving-horizon Estimation for Nonlinear Systems:
From Perfect to Imperfect Optimization}

\newif\ifuniqueAffiliation
\uniqueAffiliationtrue

\ifuniqueAffiliation 
\author{Angelo Alessandri\\
	University of Genoa (DIME)\\
	Via Opera Pia 15, 16145 Genova, Italy\\
	\texttt{angelo.alessandri@unige.it} 
}
\else
\usepackage{authblk}

\newbox{\orcid}\sbox{\orcid}{\includegraphics[scale=0.06]{orcid.pdf}} 
\author[1]{%
Angelo Alessandri\thanks{\texttt{angelo.alessandri@unige.it}}}%
}
\affil[1]{University of Genoa (DIME), Via Opera Pia 15, I-16145 Genova, Italy}
\fi

\hypersetup{
pdftitle={Robust Moving-horizon Estimation for Nonlinear Systems:
From Perfect to Imperfect Optimization},
pdfsubject={q-bio.NC, q-bio.QM},
pdfauthor={Angelo Alessandri},
pdfkeywords={robust moving-horizon estimation, incremental input/output-to-state stability, optimization, multiplicative noise.},
}

\begin{document}

\maketitle

\begin{abstract}
Robust stability of moving-horizon estimators is investigated for nonlinear
discrete-time systems that are detectable in the sense of incremental input/output-to-state stability and are affected by disturbances. The estimate of a moving-horizon 
estimator stems from the on-line solution of a least-squares minimization
problem at each time instant. 
The resulting stability guarantees depend on the optimization 
tolerance in solving such minimization problems.
Specifically, two main contributions are established: (i) the robust stability of the 
estimation error, while supposing
to solve exactly the on-line minimization problem; (ii) the practical robust stability
of the estimation error with state estimates obtained by an imperfect
minimization.
Finally, the construction of such
robust moving-horizon estimators and the performances resulting from the design
based on the theoretical findings are showcased with two numerical examples.
\end{abstract}


\section{Introduction}

We investigate the robust stability of the estimation error for moving-horizon estimators,
which result from the exact on-line minimization of a least-squares cost function by using
only a recent batch of information on the plant under monitoring. Specifically,
we will show how to ensure robustness according to
\cite{AllanRawlTeel21} by suitably discounting the effect of all the previous noises on the
current state estimate, i.e., 
in other words, by properly choosing the cost function. Moreover, it will be shown
that the estimation error is practically robustly stable with imperfect
on-line optimization, namely by stopping the optimization if the norm of the gradient of
the cost function is sufficiently small but not necessarily zero.

After the introduction of the input-to-state stability (ISS) in \cite{Sontag89},
output-to-state stability (OSS) is proposed by Sontag and Wang in \cite{SontWang97} 
as extension of detectability to nonlinear systems without inputs. Incremental
input/output-to-state stability (i-IOSS) is adopted in \cite{SontWang97}
to deal with inputs and outputs separately in general as well as it is
shown that systems admitting robust estimators must be i-IOSS. 
Unfortunately, these results do not apply to moving-horizon estimators;
only in \cite{JiRawlHuWynnDiehl16,Muller17} the robust stability of
moving-horizon estimation under the assumption of bounded disturbances was related to i-IOSS. 
Robust stability was proved without this boundedness assumption in \cite{AllanRawl21},
but under suitable conditions derived from the solution of the so-called full information
estimation (FIE) problem. Other or slightly different conditions 
were considered in \cite{KnuMull18} and later in \cite{KnuMull23,SchiMull23,Hu24},
and finally relaxed in \cite{SchiMuntwKohZeilMull23} for 
incremental uniform exponential input/output-to-state stable (i-UEIOSS) systems.

The characterization of i-IOSS in terms of Lyapunov functions is addressed
in \cite{AllanRawlTeel21} for discrete-time nonlinear
systems under quite general assumptions and in line with previous results by 
Angeli (see \cite{AngeliTAC02,AngeliTAC09}) for continuous-time
nonlinear systems. 
The findings of \cite{AllanRawlTeel21} on i-UIOSS (incremental
uniform IOSS) and i-UEIOSS  have provided the
design of moving-horizon estimators for nonlinear discrete-time systems with
a solid basis. 
First results based on this ground are presented in
\cite{AllanPhDTh20,AllanRawl21,KnuMull23,Hu24}, where the link between FIE and moving horizon
estimation is investigated. 
All the approaches to moving-horizon estimation recalled so far demand the exact
solution of the on-line minimization problems. 
In this paper, we present new results on moving-horizon estimation under imperfect
optimization, i.e., we suppose to stop the minimization without exactly finding the
minimizer of the cost in such a way as to reduce the computational burden.
In this case, practical robust stability is proved to hold. 

Reducing the on-line computational effort is a crucial issue in moving-horizon estimation
\cite{KuhlDiehKrausScholBock11,WynnVukDieh14}. Among others, in \cite{AleGagTAC17,AleGag20}
moving-horizon estimation was addressed by using only few predetermined iterations of descent methods. Accelerated gradient-based approaches to moving-horizon estimation
were investigated in \cite{GhaGhaEben23,LiuChaChoCDC23}.
Compared to all the latter, here we rely on a relaxed stopping criterion,
in that the descent algorithm is terminated if the gradient of the cost
function is sufficiently small.

The paper is organized as follows. Section \ref{sec:perfect:opt} is focused on moving
horizon estimation with exact solution of the on-line minimization
problem. 
In case of perfect optimization, we assume i-UIOSS, which is more general than i-UEIOSS.
In Section \ref{sec:imperfect:opt}, we deal with the
robust stability properties of the considered moving-horizon estimators in case
of imperfect optimization and assuming i-UEIOSS.
Section \ref{sec:sim} illustrates two case studies, while also 
including and discussing numerical results.
Finally, conclusions are given in Section~\ref{sec:concl}. 

Given the vectors
$\, v_i, v_{i+1}, \ldots , v_j $ for $i<j$, we define $\, v_i^j :=  
\left( v_i , v_{i+1} , \ldots , v_j \right) $. $\Re_{\ge 0}$ and
$\Na_{\ge k}$ denote $[0,+\infty)$ and $\{ k,k+1,\ldots\}$ with $k$ integer,
respectively.
A function $\alpha : \Re_{\ge 0}
\rightarrow \Re_{\ge 0}$ is of class $\cK$ if it is continuous, strictly increasing,
and $\alpha(0)=0$. If in addition $\lim_{s\rightarrow\infty} \alpha(s)=+\infty$,
it is of class $\cK_\infty$.
A function $\beta : \Re_{\ge 0} \times \Na_{\ge 0} 
\rightarrow \Re_{\ge 0}$ is of class $\cKL$ if $\beta(\cdot,k)$ is of class $\cK$
for any fixed $k \in \Na_{\ge 0}$ and $\beta(s,\cdot)$ is nonincreasing and
$\lim_{k\rightarrow\infty} \beta(s,k)=0$ for any fixed $s \in \Re_{\ge 0}$.
Let us denote by $\oplus$ and $\wedge$ the maximum and minimum operators, i.e.,
$a \oplus b := \max(a,b)$  and $a \wedge b := \min(a,b)$ for any $a, b \in \Re$,
respectively. 
Moreover, for the sake of brevity let us define 
$$
\dsst \Oplus_{i=1}^n a_i := \max \left( a_1, \ldots , a_n \right)
$$
for any $n \in \Na_{\ge 2}$.
The $\oplus$ operator has higher precedence than the scalar product, i.e., $c \, a
\oplus b = (ac) \oplus (bc)$ for any $a, b \in \Re$ and $c \in \Re_{\ge 0}$. For
any column vector $x \in \Re^n$, $|x| := \sqrt{x^\top x}$ denotes the Euclidean norm;
$x > 0$ means that $x_i >0$, $i=1,\ldots,n$.

\section{ Moving-horizon Estimation with Perfect Optimization}
\label{sec:perfect:opt}

In this section we will address moving-horizon estimation for
discrete-time systems described by
\begin{equation}
\label{eq:system:general}
\left\{
\begin{array}{l}
x_{t+1} =  f(x_t,w_t)  \\
y_t = h(x_t,v_t)
\end{array} \,,\, t=0,1,\ldots
\right.
\end{equation}
where  $x_t \in \Bbb X \subseteq \Re^n $ is the state vector, $y_t \in \Re^m $
is the measurement vector, $w_t \in \Bbb W \subseteq \Re^n$ is the system noise, and
$v_t \in \Bbb V \subseteq \Re^m$ is the measurement noise. 
For such systems we assume that i-IOSS holds uniformly according to the following definition, where we denote
by $x_t(\bar x,\bar w_0^{t-1})$ the state of the system \eqref{eq:system:general} at
time $t=1,2,\ldots$ with initial state $\bar x \in \Bbb X$ in $t=0$ and system
disturbances $\bar w_0, \ldots, \bar w_{t-1} \in \Bbb W$.

\medskip

\begin{definition}\label{def:i-UIOSS}
{\em System \eqref{eq:system:general} is i-UIOSS (uniformly i-IOSS\footnote{In line 
with \cite{AllanRawlTeel21}, i-IOSS is meant to be uniform w.r.t. all the external 
inputs.}) if there exist $\cKL$ functions 
$\beta_x(\cdot,\cdot), \beta_w(\cdot,\cdot)$, $\beta_y(\cdot,\cdot)$,
$\beta_v(\cdot,\cdot)$
such that
\begin{align}
&\left| x_{t}(\bar x,\bar w_0^{t-1}) - x_{t}(\tilde x,\tilde w_0^{t-1})
\right| \le \beta_x \left( \left| \bar x -  \tilde x \right| , t \right)
\oplus \Oplus_{i=0}^{t-1} \beta_w \left( | \bar w_i - \tilde w_i |
, t\!-\!1\!-\!i \right)  \notag \\
&\oplus \Oplus_{i=0}^{t-1} \beta_y \big( \big| h(x_{i}(\bar x,\bar w_0^{i-1}),\bar v_i) - h(x_{i}(\tilde x,\tilde w_0^{i-1}),\tilde v_i) 
\big| , 
t\!-\!1\!-\!i \big) \oplus \Oplus_{i=0}^{t-1} \beta_v \left( | \bar v_i - \tilde v_i | 
, t\!-\!1\!-\!i \right) \,,\, t=0,1,\ldots
\label{eq:def:i-UIOSS}
\end{align}
for all $\bar x, \tilde x \in \Bbb X$, $\bar w_t, \tilde w_t \in \Bbb W$, 
$\bar v_t, \tilde v_t \in \Bbb V$.
}
\end{definition}

Definition \ref{def:i-UIOSS} was proposed in 
\cite[see Definition 1, p. 3]{KnuMull23}) and turns out to be quite general 
as compared to what is available in the literature since it allows to deal with
multiplicative measurement noises, as it will be shown in 
Section \ref{sec:example:2}.

\medskip

\begin{remark}
In principle, there are no difficulties in developing what is presented
in the following by using a system representation that includes
some known input $u_t \in \Bbb U \subseteq \Re^p$ in the system dynamics, i.e.,
using $x_{t+1} =  f(x_t,u_t,w_t)$ instead of $x_{t+1} =  f(x_t,w_t)$.
Unfortunately, this would complicate the notation, thus we prefer to
treat the problems in a simpler form without $u_t$.
\end{remark}


As shown in \cite{AllanRawlTeel21}, it is necessary for a system to be i-IUOSS
for admitting a robust estimator, where, as estimator, we mean a generic
input/output mapping between the current available information up to
$t$ and the estimate of $x_t$ denoted by $\hat x_{t|t} \in \Bbb X$,
starting just with some ``a priori'' prediction of $x_0$ denoted by
$\bar x_0 \in \Bbb X$.


\begin{definition}\label{def:robust:estimator}
{\em An estimator for system \eqref{eq:system:general} 
with $\hat x_{t|t} \in \Bbb X$ as estimate of $x_t$ at time $t$
is robustly stable if there exist 
$\cKL$ functions $\rho_x(\cdot,\cdot), \rho_w(\cdot,\cdot), \rho_v(\cdot,\cdot)$
such that
\begin{align}
&\left| x_{t} - \hat x_{t|t} \right| \le \rho_x \left( \left| x_{0}
- \bar x_{0} \right| ,t \right) \oplus \Oplus_{i=0}^{t-1} \rho_w \left(
\left| w_i \right| , t\!-\!1\!-\!i \right) 
\oplus \Oplus_{i=0}^{t-1} \rho_v \left(
\left| v_i \right| , t\!-\!1\!-\!i \right) \label{eq:def:robust:estimator}
\end{align}
for all $t=0,1,\ldots$.
}
\end{definition} 

\medskip

Definition \ref{def:i-UIOSS} is equivalent to detectability. For detectable linear
discrete-time systems, it is easily shown the robust stability of the estimation error w.r.t
noises by using summation instead of maximization in Definition 
\ref{def:robust:estimator} (see, e.g., \cite[pp. 3020-3021]{AllanRawlTeel21}).

We consider the problem of estimating $x_t$ at time $t$ according to the moving-horizon
paradigm, i.e., by using the batch of the last $t \wedge \! N$ measurements $y_{t-t \wedge \! N} , \ldots , y_{t-1}$. 
At any $t$ we obtain the estimate $\hat x_{t|t} \in \Bbb X$ of $x_t$
by minimizing a cost function where the deviation from some ``prediction''
\footnote{The prediction summarizes the previous knowledge on the state
variables at time $t-t \wedge \! N$, i.e., at the beginning of the moving window. From $t=1$ to $t=N$ it
is kept equal to an ``a priori'' constant vector $\bar x_0 \in \Bbb X$.
}
of the state $x_{t-t \wedge \! N} $ (denoted by $\bar x_{t-t \wedge N}$) and
the fitting of $y_{t-t \wedge \! N} , \ldots , y_{t-1}$ are penalized as follows
\begin{align}
&J_t^{t \wedge \! N} \left(\hat x_{t-t \wedge \! N} , \hat v_{t-t \wedge \! N}^{t-1} ,
\hat w_{t-t \wedge \! N}^{t-1}  \right) =  \beta_x ( 2 \, |\hat x_{t-t \wedge \! N}
- \bar x_{t-t \wedge \! N}| , {t \wedge \! N} ) \oplus \Oplus_{i=t-t \wedge \! N}^{t-1} \, \beta_w \left(
2 \left| \hat w_{i} \right| , t\!-\!1\!-\!i \right) \notag \\
&\oplus \Oplus_{i=t-t \wedge \! N}^{t-1} \, \beta_y \left( \left| y_i -
h(\hat x_{i},\hat v_{i}) \right| , t\!-\!1\!-\!i \right) 
\oplus \Oplus_{i=t-t \wedge \! N}^{t-1} \, \beta_v \left(
2 \left| \hat v_{i} \right| , t\!-\!1\!-\!i \right) \,,\, t=1,2,\ldots
\label{eq:cost:general}
\end{align}
where $\beta_x(\cdot,\cdot), \beta_w(\cdot,\cdot),
\beta_y(\cdot,\cdot), \beta_v(\cdot,\cdot)$ are the same $\cKL$ functions
involved in the definition of i-UIOSS for system \eqref{eq:system:general}
and the constraints
$$
\hat x_{i+1} =  f ( \hat x_{i} , \hat w_{i} )
\,,\, i=t-t \wedge \! N,\ldots,t-1 
$$
are implicitly considered in \eqref{eq:cost:general}; moreover, we rely on the
predictions given by 
\begin{equation}
\bar x_{t-t \wedge N} =  \left\{
\begin{array}{ll}
\bar x_0           & t=1,\ldots,N \\
\hat x_{t-N|t-N} & t=N+1,N+2,\ldots \,.
\end{array}
\right.
\label{eq:prediction}
\end{equation}

Since the cost is continuous, minimizers of \eqref{eq:cost:general}
exist under the usual assumptions required by the Weierstrass' theorem, i.e.,
lower semicontinuity of the cost function and compactness or closeness of
the domain (coercivity of the cost is required too in case closeness holds
but compactness does not)
\cite[Proposition A.8, p. 669]{Berts99}. In this respect, $f(\cdot,\cdot)$ and
$h(\cdot,\cdot)$ are assumed to be smooth enough as well. 

We denote by MHE$_N$ a generic moving-horizon estimator, which provides an estimate
$\hat x_{t|t}$ of $x_t$ at time $t$ as the result of minimization of \eqref{eq:cost:general}, together with \eqref{eq:prediction}. Thus, at each $t=1,2,\ldots$ we have to solve
\begin{align}
\left( \hat x_{t-t\wedge N|t} , \hat v_{t-t\wedge N|t}^{t-1} , 
\hat w_{t-t\wedge N|t}^{t-1} \right) &\in
\dsst \!\!\!\! \ARGMIN_{\begin{array}{c} \hat x_{t-t\wedge N} \in \Bbb X , \hat v_{t-t\wedge N}^{t-1}
\! \in \\ \Bbb V^{t\wedge N}, \hat w_{t-t\wedge N}^{t-1} \in \Bbb W^{t\wedge N} \end{array}}
\!\!\!\!\!\! 
J_t^{t \wedge \! N} \left( \hat x_{t-t\wedge N} , \hat v_{t-t\wedge N}^{t-1}, 
\hat w_{t-t\wedge N}^{t-1} \right)
\label{eq:mhe:general}
\end{align}
and we get
$\hat x_{t|t}$ by means of
$
\hat x_{i+1|t} =  f ( \hat x_{i|t} , \hat w_{i|t} )
\,,\, i=t-t \wedge \! N,\ldots,t-1 \,.
$

\medskip

\begin{theorem}\label{theo:RS:general}
{\em If system \eqref{eq:system:general} is i-UIOSS and there exist $\cKL$ functions 
$\alpha_x(\cdot,\cdot), \alpha_v(\cdot,\cdot)$, $\alpha_w(\cdot,\cdot)$ and
$N \in \Na_{\ge 1}$ such that 
\begin{subequations}\label{eq:cond:theo:RS:general:alpha:beta:x}
\begin{align}
&\beta_x \left( 4 \beta_x(2s,k) , N \right) \le \alpha_x(s,k+N)
\label{eq:cond:theo:RS:general:alpha:beta:1:x} \\
&\beta_x \left( 2 \alpha_x(s,k) , N \right) \le \alpha_x(s,k+N) 
\label{eq:cond:theo:RS:general:alpha:beta:2:x}  \\[-1.0cm] \notag
\end{align}
\end{subequations}
\begin{subequations}\label{eq:cond:theo:RS:general:alpha:beta:w}
\begin{align}
&\beta_x \left( 4 \beta_w(2s,k) , N \right) \le \alpha_w(s,k+N)
\label{eq:cond:theo:RS:general:alpha:beta:1:w}  \\
&\beta_x \left( 2 \alpha_w(s,k) , N \right) \le \alpha_w(s,k+N)
\label{eq:cond:theo:RS:general:alpha:beta:2:w}   \\[-1.0cm] 
\notag
\end{align}
\end{subequations}
\begin{subequations}\label{eq:cond:theo:RS:general:alpha:beta:v}
\begin{align}
&\beta_x \left( 4 \beta_v(2s,k) , N \right) \le \alpha_v(s,k+N)
\label{eq:cond:theo:RS:general:alpha:beta:1:v}  \\
&\beta_x \left( 2 \alpha_v(s,k) , N \right) \le \alpha_v(s,k+N)
\label{eq:cond:theo:RS:general:alpha:beta:2:v}
\end{align}
\end{subequations}
for all $(s,k) \in [0,+\infty) \times \Na_{\ge 0}$, the MHE$_N$
resulting from \eqref{eq:mhe:general} is robustly stable.
}
\end{theorem}

\noindent{\em Proof}. From the definition of minimum of the cost function, i.e.,
\begin{align}
J_t^{t \wedge \! N} \left(\hat x_{t-t \wedge \! N|t} , \hat v_{t-t \wedge \! N|t}^{t-1} ,
\hat w_{t-t \wedge \! N|t}^{t-1}  \right) 
\le  J_t^{t \wedge \! N} \left( x_{t-t \wedge \! N} , v_{t-t \wedge \! N}^{t-1} , w_{t-t \wedge 
\! N}^{t-1} \right) \,,
\notag
\end{align}
for any $t=1,2,\ldots$ we obtain
\begin{align}
&\beta_x \left( 2 \left| \hat x_{t-t \wedge \! N|t}
- \bar x_{t-t \wedge \! N} \right| , t \wedge \! N \right) \oplus \Oplus_{i=t-t \wedge \! N}^{t-1} \beta_y
\big( \big| y_i - 
h\left(\hat x_{i|t},\hat v_{i|t}\right) \big|  , t\!-\!1\!-\!i \big) 
\oplus \Oplus_{i=t-t \wedge \! N}^{t-1} \hat \beta_v \left(
2 \left| \hat v_{i|t} \right| , t\!-\!1\!-\!i \right) \notag \\
&\oplus \Oplus_{i=t-t \wedge \! N}^{t-1} \beta_w \left(
2 \left| \hat w_{i|t} \right| , t\!-\!1\!-\!i \right)\notag \\
&\le \beta_x ( 2 |x_{t-t \wedge \! N} - \bar x_{t-t \wedge \! N}| , t \wedge \! N ) 
\oplus \Oplus_{i=t-t \wedge \! N}^{t-1} \beta_y \big(
\underbrace{
\left| y_i - h(x_{i},v_{i}) \right|}_{=0} , t\!-\!1\!-\!i \big)
\oplus \Oplus_{i=t-t \wedge \! N}^{t-1} \beta_v \left(
2 \left| v_{i} \right| , t\!-\!1\!-\!i \right) \notag \\[-0.2cm]
&\oplus \Oplus_{i=t-t \wedge \! N}^{t-1} \beta_w \left(
2 \left| w_{i} \right| , t\!-\!1\!-\!i \right) \,.
\label{eq:theo:RS:general:bound:min}
\end{align}

Let us focus on three types of inequality to manipulate inequality
\eqref{eq:theo:RS:general:bound:min}. From
\begin{align}
\left| x_{t-t \wedge \! N} - \hat x_{t-t \wedge \! N|t} \right| &\le
\left| x_{t-t \wedge \! N} - \bar x_{t-t \wedge \! N} \right| 
+ \big| \hat x_{t-t \wedge \! N|t} - \bar x_{t-t \wedge \! N} \big| \notag
\end{align}
by using $\rho(a+b) \le \rho(2a) + \rho(2b), \forall a,b \in \Re_{\ge 0}$
with $\rho(\cdot)$ being any $\cK$ function
\cite[Equation (12), p. 438]{Sontag89} it follows that
\begin{align}
&\beta_x \left( \left| x_{t-t \wedge \! N} - \hat x_{t-t \wedge \! N|t} \right| ,
t \wedge \! N \right) - \beta_x ( 2 | x_{t-t \wedge \! N} 
-  \bar x_{t-t \wedge \! N} | ,
t \wedge \! N ) \le \beta_x \left( 2 \left| \hat x_{t-t \wedge \! N|t} - 
\bar x_{t-t \wedge \! N} \right|
, t \wedge \! N \right) \,. \label{eq:cond:theo:RS:general:bound:1}
\end{align}
Likewise, from $\left|v_{i}- \hat v_{i|t} \right| \le |v_{i}| +
\left|\hat v_{i|t}\right|$ and $\left|w_{i}- \hat w_{i|t}\right| \le
|w_{i}| + \left|\hat w_{i|t}\right|$
we obtain
\begin{align}
&\beta_v \left( \left| v_{i}- \hat v_{i|t} \right| ,
t-1-i \right) - \beta_v \left( 2 |v_{i}| , t-1-i\right) 
\le \hat \beta_v \left( 2 \left|\hat v_{i|t}\right|
, t-1-i \right) \label{eq:cond:theo:RS:general:bound:2} \\
&\beta_w \left( \left| w_{i}- \hat w_{i|t} \right| ,
t-1-i \right) - \beta_w \left( 2 |w_{i}| , t-1-i\right) 
\le \beta_w \left( 2 \left|\hat w_{i|t}\right|
, t-1-i \right) \label{eq:cond:theo:RS:general:bound:3}
\end{align}
for $i\!=\!t\!-\!t \wedge \! N,\ldots,t\!-\!1$. 

We will rely on
\eqref{eq:cond:theo:RS:general:bound:1}, \eqref{eq:cond:theo:RS:general:bound:2}, 
and \eqref{eq:cond:theo:RS:general:bound:3} to deal with
\eqref{eq:theo:RS:general:bound:min}. Specifically, by using the upper bound
$
a + a \oplus b \le 2 \, a \oplus b \,,\, \forall a,b \in \Re
$,
and \eqref{eq:cond:theo:RS:general:bound:1}, from
\eqref{eq:theo:RS:general:bound:min} it follows that
\begin{align}
&\beta_x \left( \left| x_{t-t \wedge \! N} - \hat x_{t-t \wedge \! N|t} \right| ,
t \wedge \! N \right)
\le 2 \, \beta_ x \big( 2 | x_{t-t \wedge \! N} - \bar x_{t-t \wedge \! N} | , t \wedge \! N \big) \oplus \Oplus_{i=t-t \wedge \! N}^{t-1} \beta_v \left(
2 \left| v_{i} \right| , t\!-\!1\!-\!i \right) \notag \\
&\oplus \Oplus_{i=t-t \wedge \! N}^{t-1} \beta_w \left(
2 \left| w_{i} \right| , t\!-\!1\!-\!i \right) \,.
\label{eq:theo:RS:general:bound:min:after:1}
\end{align}
Similar bounds are established from \eqref{eq:theo:RS:general:bound:min}
by using \eqref{eq:cond:theo:RS:general:bound:2}, i.e.,
\begin{align}
&\beta_v \left( \left| v_{i}- \hat v_{i|t} \right| ,
t \wedge \! N \right) \le 2 \, \beta_x \big( 2 | x_{t-t \wedge \! N} -
\bar x_{t-t \wedge \! N} | , N \big) 
\oplus \Oplus_{i=t-t \wedge \! N}^{t-1} \beta_v \left(
2 \left| v_{i} \right| , t\!-\!1\!-\!i \right) \notag \\
&\oplus \Oplus_{i=t-t \wedge \! N}^{t-1} \beta_w\left(
2 \left| w_{i} \right| , t\!-\!1\!-\!i \right) \,,\, i\!=\!t\!-\!t \wedge \! N,
\ldots,t\!-\!1 \,,
\label{eq:theo:RS:general:bound:min:after:2}
\end{align}
and \eqref{eq:cond:theo:RS:general:bound:3}, i.e.,
\begin{align}
&\beta_w \left( \left| w_{i}- \hat w_{i|t} \right| ,
t \wedge \! N \right) \le 2 \, \beta_x \big( | x_{t-t \wedge \!N}
- \bar x_{t-t \wedge \! N} | , N \big) \oplus \Oplus_{i=t-t \wedge \! N}^{t-1} 
\beta_v \left( 2 \left| v_{i} \right| , t\!-\!1\!-\!i \right) \notag \\
&\oplus \Oplus_{i=t-t \wedge \! N}^{t-1} \beta_w \left(
2 \left| w_{i} \right| , t\!-\!1\!-\!i \right) \,,\, i\!=\!t\!-\!t \wedge \! N,
\ldots,t-1 .
\label{eq:theo:RS:general:bound:min:after:3}
\end{align}
Moreover, \eqref{eq:theo:RS:general:bound:min} yields
\begin{align}
\!\!\!\!\Oplus_{i=t-t \wedge \! N}^{t-1} \beta_y \big( \big| h(x_i,v_i) -
h\left(\hat x_{i|t},\hat v_{i|t}\right) \big|  , t\!-\!1\!-\!i \big) &\le \beta_x \big( \big| x_{t-t \wedge \!N} - \bar x_{t-t \wedge \!N} \big| , N \big)
\oplus \Oplus_{i=t-t \wedge \! N}^{t-1} \beta_v \left(
2 \left| v_{i} \right| , t\!-\!1\!-\!i \right) \notag \\
&\oplus \Oplus_{i=t-t \wedge \! N}^{t-1} \beta_w \left(
2 \left| w_{i} \right| , t\!-\!1\!-\!i \right) \notag \\
&\le 2 \beta_x \big( 2 \big| x_{t-t \wedge \!N} - \bar x_{t-t \wedge \!N} \big| , N \big) 
\oplus \Oplus_{i=t-t \wedge \! N}^{t-1} \beta_v \left(
2 \left| v_{i} \right| , t\!-\!1\!-\!i \right) \notag \\
&\oplus \Oplus_{i=t-t \wedge \! N}^{t-1} \beta_w \left(
2 \left| w_{i} \right| , t\!-\!1\!-\!i \right) \,,\, i\!=\!t\!-\!t \wedge \! N,
\ldots,t-1 .
\label{eq:theo:RS:general:bound:min:after:4}
\end{align}
Bringing \eqref{eq:theo:RS:general:bound:min:after:1}-\eqref{eq:theo:RS:general:bound:min:after:4} all together, we obtain 
\begin{align}
&\beta_x \left( \left| x_{t-t \wedge \! N} - \hat x_{t-t \wedge \! N|t} \right| ,
t \wedge \! N \right) 
\oplus \Oplus_{i=t-t \wedge \! N}^{t-1}
\beta_y \big( \big| h(x_i,v_i) - h\left(\hat x_{i|t},\hat v_{i|t}\right) \big|  , t\!-\!1\!-\!i \big) \notag \\
&\oplus \Oplus_{i=t-t \wedge \! N}^{t-1} \beta_v \left( 
\left| v_{i}- \hat v_{i|t} \right| , t\!-\!1\!-\!i \right)
\oplus \Oplus_{i=t-t \wedge \! N}^{t-1} \beta_w
\left( \left| w_{i}- \hat w_{i|t} \right| ,
t\!-\!1\!-\!i \right) \notag \\
&\le 2 \, \beta_x \big( 2 \, |x_{t-t \wedge \! N} - \bar x_{t-t \wedge \! N}| , N \big)
\oplus \Oplus_{i=t-t \wedge \! N}^{t-1} \beta_v \left(
2 \left| v_{i} \right| , t\!-\!1\!-\!i \right) 
\oplus \Oplus_{i=t-t \wedge \! N}^{t-1} \beta_w \left(
2 \left| w_{i} \right| , t\!-\!1\!-\!i \right) \notag
\end{align}
and thus
\begin{align}
&\left| x_{t} - \hat x_{t|t} \right| \le 2 \, \beta_x \big( 2 \, 
| x_{t-t \wedge \! N}  - \bar x_{t-t \wedge \! N} | , t \wedge \! N \big) \oplus \Oplus_{i=t-t \wedge \! N}^{t-1} \beta_v \left(
2 \left| v_{i} \right| , t\!-\!1\!-\!i \right) \notag \\
&\oplus \Oplus_{i=t-t \wedge \! N}^{t-1} \beta_w \left( 2
\left| w_{i} \right| , t\!-\!1\!-\!i \right) \,,\, t=1,2,\ldots \,,
\label{eq:theo:RS:general:bound:fundam}
\end{align}
since system \eqref{eq:system:general} is assumed to be i-UIOSS. 

In the following, first of all we analyze the bounds derived from
\eqref{eq:theo:RS:general:bound:fundam} when (i) $t=1,\ldots,N$,
(ii) $t=N+1,\ldots,2N$, and (iii) $t \ge 2N+1$. Later, we will
combine the previous results to prove robust stability
in the sense of Definition \ref{def:robust:estimator}.

Concerning (i), from \eqref{eq:theo:RS:general:bound:fundam}
it is straightforward to get
\begin{align}
&\left| x_{t} - \hat x_{t|t} \right| \le 2 \, \beta_x \big( 2 | 
x_0  - \bar x_0 | , t \big)
\oplus \Oplus_{i=0}^{t-1} \beta_v \left(
2 \left| v_{i} \right| , t\!-\!1\!-\!i \right) 
\oplus \Oplus_{i=0}^{t-1} \beta_w \left(
2 \left| w_{i} \right| , t\!-\!1\!-\!i \right) \,,\, t=1,\ldots,N \,.
\label{eq:theo:RS:general:bound:fundam:t:1:N}
\end{align}
Let us consider (ii). From \eqref{eq:theo:RS:general:bound:fundam} for 
$t=N+1$ we obtain
\begin{align}
&\left| x_{N+1} - \hat x_{N+1|N+1} \right| \le 2 \, \beta_x
\big( 2 \, | x_1 - \hat x_{1|1} | , N \big)
\oplus \Oplus_{i=1}^{N} \beta_v \left(
2 \left| v_{i} \right| , N \!-\!i \right) 
\oplus \Oplus_{i=1}^{N} \beta_w \left(
2 \left| w_{i} \right| , N\!-\!i \right)
\label{eq:theo:RS:general:bound:fundam:t=N+1}
\end{align}
since $\bar x_1 = \hat x_{1|1}$ and, using
\eqref{eq:theo:RS:general:bound:fundam:t:1:N} for $t=1$, it follows that
\begin{align}
\left| x_{N+1} - \hat x_{N+1|N+1} \right| &\le 2 \, \beta_x
\big( 4 \beta_x \big( 2 | x_0  - \bar x_0 | , 1 \big) 
\oplus \beta_v (2 |v_{0}| , 0) \oplus \beta_w ( 2 |w_{0}| , 0 )
, N \big) \oplus \Oplus_{i=1}^{N} \beta_v \left( \left| v_{i} \right| , N \!-\!i \right) 
\notag \\
&\oplus \Oplus_{i=1}^{N} \beta_w \left(
\left| w_{i} \right| , N\!-\!i \right) \notag \\
&= 2 \, \beta_x \big( 4 \beta_x \big( 2 | x_0  - \bar x_0 | , 1 \big) , N \big)
\oplus \beta_x \big( \beta_v (2 |v_{0}| , 0) , N \big) 
\oplus \beta_x \big( \beta_w (2 |w_{0}| , 0) , N \big) \notag \\
&\oplus \Oplus_{i=1}^{N} \beta_v \left(
2 \left| v_{i} \right| , N \!-\!i \right) \oplus
\Oplus_{i=1}^{N} \beta_w \left( 2 \left| w_{i} \right| , N\!-\!i \right) \notag \\
&\le 2 \, \alpha_x \big( |x_0  - \bar x_0| , N+1 \big)  
\oplus \alpha_v( |v_{0}| , N) 
\oplus \Oplus_{i=1}^{N} \beta_v \left(
2 \left| v_{i} \right| , N \!-\!i \right) \oplus \alpha_w( |w_{0}| , N)
\notag \\
&\oplus \Oplus_{i=1}^{N} \beta_w \left( 2 \left| w_{i} \right| , N\!-\!i \right)
\label{eq:theo:RS:general:bound:fundam:t:N+1:only}
\end{align}
by using \eqref{eq:cond:theo:RS:general:alpha:beta:1:x}, 
\eqref{eq:cond:theo:RS:general:alpha:beta:1:v}, and
\eqref{eq:cond:theo:RS:general:alpha:beta:1:w}. Proceeding similarly for
$t=N+2,\ldots,2N$, we get
\begin{align}
\left| x_{t} - \hat x_{t|t} \right| &\le 2 \, \alpha_x \big( 2 | 
x_0  - \bar x_0 | , t \big) 
\oplus \Oplus_{i=0}^{t-N-1} \alpha_v \left(
\left| v_{i} \right| , t\!-\!1\!-\!i \right) 
\oplus \Oplus_{i=t-N}^{t-1} \beta_v \left(
2 \left| v_{i} \right| , t\!-\!1\!-\!i \right) \notag \\
&\oplus \Oplus_{i=0}^{t-N-1} \alpha_w \left(
\left| w_{i} \right| , t\!-\!1\!-\!i \right)
\oplus \Oplus_{i=t-N}^{t-1} \beta_w \left(
2 \left| w_{i} \right| , t\!-\!1\!-\!i \right) \,,\, t=N+1,\ldots,2N \,.
\label{eq:theo:RS:general:bound:fundam:t:N+1:2N}
\end{align}
In case of (iii), for $t=2N+1$ we need to use both \eqref{eq:theo:RS:general:bound:fundam}
and \eqref{eq:theo:RS:general:bound:fundam:t:N+1:2N}. It follows that
\begin{align}
\left| x_{2N+1} - \hat x_{2N+1|2N+1} \right| &\le 2 \, \beta_x \big( 2 \, 
| x_{N+1}  - \hat x_{N+1|N+1} | , N \big) 
\oplus \Oplus_{i=N+1}^{2N} \beta_v \left(
2 \left| v_{i} \right| , 2N\!-\!i \right) \notag \\
&\oplus \Oplus_{i=N+1}^{2N} \beta_w \left( 2
\left| w_{i} \right| , 2N\!-\!i \right) \notag
\end{align}
and thus by using \eqref{eq:theo:RS:general:bound:fundam:t:N+1:only}
\begin{align}
\left| x_{2N+1} - \hat x_{2N+1|2N+1} \right| &\le 2 \, \beta_x \Big( 2 \,
\alpha_x \big( |x_0  - \bar x_0| , N+1 \big) 
\oplus \alpha_v( |v_{0}| , N) 
\oplus \Oplus_{i=1}^{N} \beta_v \left( 2 \left| v_{i} \right| , N \!-\!i \right)
\oplus \alpha_w( |w_{0}| , N) \notag \\
&\oplus \Oplus_{i=1}^{N} \beta_w \left( 2 \left| w_{i}
\right| , N\!-\!i \right) , N \Big) \oplus \Oplus_{i=N+1}^{2N} \beta_v \left(
2 \left| v_{i} \right| , 2N\!-\!i \right) \notag \\
&\oplus \Oplus_{i=N+1}^{2N} \beta_w \left( 2
\left| w_{i} \right| , 2N\!-\!i \right) \notag \\
&\le 2 \, \beta_x \Big( 2 \, \alpha_x \big( |x_0  - \bar x_0| , N+1 \big)
, N \Big) 
\oplus\beta_x \Big( 2 \, \alpha_v( |v_{0}| , N) 
, N \Big) 
\oplus
\beta_x \Big( 2 \, \alpha_w( |w_{0}| , N) 
, N \Big) \notag \\
&\oplus
\Oplus_{i=1}^{N} \beta_x \Big(
\beta_v \left( 2 \left| v_{i} \right| , N \!-\!i \right)
, N \Big) 
\oplus \Oplus_{i=1}^{N} \beta_x \Big(
\beta_w \left( 2 \left| w_{i} \right| , N \!-\!i \right)
, N \Big) \notag \\
&\oplus
\Oplus_{i=N+1}^{2N} \beta_v \left(
2 \left| v_{i} \right| , 2N\!-\!i \right) 
\oplus \Oplus_{i=N+1}^{2N} \beta_w \left( 2
\left| w_{i} \right| , 2N\!-\!i \right) \,.
\label{eq:theo:RS:general:bound:fundam:t:only:2N+1}
\end{align}
Since \eqref{eq:cond:theo:RS:general:alpha:beta:2:x},
\eqref{eq:cond:theo:RS:general:alpha:beta:2:v}, and
\eqref{eq:cond:theo:RS:general:alpha:beta:2:w}
yield $\beta_x \Big( 2 \,
\alpha_x \big( |x_0  - \bar x_0| , N+1 \big) , N \Big) \le \alpha_x
\big( |x_0  - \bar x_0| , 2N+1 \big)$, $\beta_x \Big( 2 \, 
\alpha_v( |v_{0}| , N)  , N \Big) \le \alpha_v ( |v_{0}| , 2N)$,
and $\beta_x \Big( 2 \, 
\alpha_w( |w_{0}| , N)  , N \Big) \le \alpha_w ( |w_{0}| , 2N)$,
respectively, as well as 
\begin{align}
&\beta_x \Big( \beta_w \left( 2 |w_{i}| , N \!-\!i \right)
, N \Big) \le \beta_x \Big( 4 \beta_w \left( 2 |w_{i}| , 
N \!-\!i \right) , N \Big) 
\le \alpha_w \left( |w_{i}| , 2N \!-\!i \right) \notag \\
&\beta_x \Big( \beta_v \left( 2 \left| v_{i} \right| , N \!-\!i \right)
, N \Big) \le \beta_x \Big( 4 \beta_v \left( 2 \left| v_{i} \right| , 
N \!-\!i \right) , N \Big) 
\le \alpha_v \left( |v_{i}| , 2N \!-\!i \right) \notag 
\end{align}
by using  \eqref{eq:cond:theo:RS:general:alpha:beta:1:w}
and \eqref{eq:cond:theo:RS:general:alpha:beta:1:v},
respectively, from \eqref{eq:theo:RS:general:bound:fundam:t:only:2N+1}
it follows that
\begin{align}
\left| x_{2N+1} - \hat x_{2N+1|2N+1} \right| &\le 2 \, \alpha_x
\big( |x_0  - \bar x_0| , 2N+1 \big)
\oplus \Oplus_{i=0}^{N} \alpha_v(|v_{i}| , 2N\!-\!i) \oplus \Oplus_{i=N+1}^{2N}
\beta_v(2 |v_{i}| , 2N\!-\!i) \notag \\
&\oplus \Oplus_{i=0}^{N} \alpha_w(|w_{i}| , 2N\!-\!i) \oplus \Oplus_{i=N+1}^{2N}
\beta_w(2 |w_{i}| , 2N\!-\!i) \,, \notag 
\end{align}
which coincides with \eqref{eq:theo:RS:general:bound:fundam:t:N+1:2N}
for $t=2N+1$. In so doing for any $t \ge  2N+2$, we get that 
\eqref{eq:theo:RS:general:bound:fundam:t:N+1:2N} holds for all $t \ge N+1$.
Therefore, from combining \eqref{eq:theo:RS:general:bound:fundam:t:1:N}
and \eqref{eq:theo:RS:general:bound:fundam:t:N+1:2N} (holding for all $t 
\ge N+1$) robust stability is established for estimator MHE$_N$
in the sense of  Definition \ref{def:robust:estimator} by taking 
$\rho_x(s,t) := 2 \, \beta_x(2s,t) \oplus \alpha_x(s,t) , \rho_v(s,t)
:= 2 \, \beta_v(2s,t) \oplus \alpha_v(s,t) , \rho_w(s,t) := 2 \, 
\beta_w(2s,t) \oplus \alpha_w(s,t)$.  \qed

\medskip

\begin{remark}
Theorem \ref{theo:RS:general} guarantees robust stability
if the minimization problem is solved exactly,  which corresponds to the choice of
$A=1$ in \cite[formulas (11), p. 4 and (22), p. 5]{KnuMull23}. Indeed, 
the satisfaction of \cite[(11), p. 4]{KnuMull23} 
is nontrivial if $A>1$, which entails some suboptimality margin but, to be
applied, demands the knowledge of the true state and disturbances.
Moreover, in \cite{KnuMull23,Hu24} additional conditions concerning the
FIE problem are required such as \cite[(23)-(25) and Theorem 14, p. 6]{KnuMull23}
and \cite[Assumption 2, p. 4 and Assumption 3, p. 6]{Hu24}. By contrast, conditions 
\eqref{eq:cond:theo:RS:general:alpha:beta:1:x}-\eqref{eq:cond:theo:RS:general:alpha:beta:2:v}
in Theorem \ref{theo:RS:general} are simple and can be rigorously checked, as shown in the case
study of Section \ref{sec:example:1}. 
\end{remark}

\medskip

Theorems \ref{theo:RS:general} 
demands to minimize the cost function perfectly. In order to overcome this limitation,
we will address moving-horizon estimation
for a class of systems with additive disturbances and a quadratic cost
function with exponential discount. More specifically, consider 
\begin{equation}
\label{eq:system}
\left\{
\begin{array}{l}
x_{t+1} =  f(x_t) + w_t  \\
y_t = h(x_t) + v_t
\end{array} \,,\, t=0,1,\ldots
\right.
\end{equation}
and assume that incremental uniform exponential input/output-to-state stability
(i-UEIOSS) holds for system \eqref{eq:system}
according to the following definition.

\medskip

\begin{definition}\label{def:i-UEIOSS}
{\em System \eqref{eq:system} is i-UEIOSS if there exist $\eta \in (0,1)$ and $c_x,
c_v, c_w > 0$
\begin{align}
\left| x_{t}(\bar x,\bar w_0^{t-1}) - x_{t}(\tilde x,\tilde w_0^{t-1})
\right|^2 &\le c_x \left| \bar x -  \tilde x \right|^2 \eta^t 
+  c_v \! \sum_{i=0}^{t-1} \! \eta^{t-1-i} \! \left| h(x_{i}(\bar x,\bar w_0^{i-1}))
\!+\! \bar v_i \!-\! h(x_{i}(\tilde x,\tilde w_0^{i-1})) \!-\! 
\tilde v_i  \right|^2 \notag \\
&+ c_w \! \sum_{i=0}^{t-1} \! \eta^{t-1-i} | \bar w_i - \tilde w_i |^2 \,,\,
t=1,2,\ldots \label{eq:def:i-UEIOSS:bound}
\end{align}
for all $\bar x, \tilde x \in \Bbb X$, $\bar v_t, \tilde v_t \in \Bbb V$, and
$\bar w_t, \tilde w_t \in \Bbb W$.
}
\end{definition}

\medskip

\begin{definition}\label{def:robust:stab:and:practical}
{\em An estimator for system \eqref{eq:system} 
with $\hat x_{t|t}$ as estimate of $x_t$ at time $t$
is exponentially robustly stable if
\begin{align}
\left| x_{t} - \hat x_{t|t} \right|^2 &\le \alpha \left| x_{0}
- \bar x_{0} \right|^2 \lambda^t + \beta \sum_{i=0}^{t-1} \lambda^{t-1-i} |v_i|^2
+ \gamma \sum_{i=0}^{t-1} \lambda^{t-1-i} |w_i|^2 \,,\,
t=1,2,\ldots \label{eq:def:exp:robust:est}
\end{align}
for some $\lambda \in (0,1)$, where $\alpha, \beta, \gamma >0$, and $\varepsilon$-practically exponentially robustly
stable for some $\varepsilon>0$ if
\begin{align}
\left| x_{t} - \hat x_{t|t} \right|^2 &\le \alpha \left| x_{0}
- \bar x_{0} \right|^2 \lambda^t + \beta \sum_{i=0}^{t-1} \lambda^{t-1-i} |v_i|^2
+ \gamma \sum_{i=0}^{t-1} \lambda^{t-1-i} |w_i|^2 + \varepsilon^2 \,,\,
t=1,2,\ldots \label{eq:def:pract:exp:robust:est}
\end{align}
where again $\alpha, \beta, \gamma >0$ and $\lambda \in (0,1)$.
}
\end{definition}

\medskip

Definitions \ref{def:i-UEIOSS} and \ref{def:robust:stab:and:practical}
are formulated by using squares of norm instead of norms
since they are equivalent as $\sqrt{a+b} \le \sqrt{a} + \sqrt{b}$,
for all $a, b \ge 0$. This allows for some simplification without loss of
generality when manipulating the inequalities necessary to prove
the next findings.

From now on we suppose to know the i-UEIOSS discount factor $\eta \in (0,1)$ or at least some upper
bound of $\eta$ just strictly less than one.\footnote{In principle,
a more general treatment can rely on the use of a cost function
\begin{align}
J_t^{t \wedge \! N} \left(\hat x_{t-t \wedge \! N}, 
\hat w_{t-t \wedge \! N}^{t-1} \right) =
\mu \, \left| \hat x_{t-t \wedge \! N} - \bar x_{t-t \wedge \! N} \right|^2 \bar \eta^{t \wedge \! N}
+ \nu \!\! \sum_{i=t-{t \wedge \! N}}^{t-1} \bar \eta^{t-1-i} \left| y_i -
h(\hat x_{i}) \right|^2
+ \omega \!\! \sum_{i=t-{t \wedge \! N}}^{t-1} \bar \eta^{t-1-i} \left| \hat w_{i}
\right|^2 \notag
\end{align}
with $\bar \eta \in [\eta,1)$, i.e.,  $\bar \eta$ being an upper bound of $\eta$
in $(0,1)$.}
Therefore, we consider the MHE$_N$ providing the estimate $\hat x_{t|t}
\in \Bbb X$ of  $x_t$, obtained by minimizing the quadratic cost 
\begin{align}
J_t^{t \wedge \! N} \left(\hat x_{t-t \wedge \! N}, 
\hat w_{t-t \wedge \! N}^{t-1} \right) =
\mu \, \left| \hat x_{t-t \wedge \! N} - \bar x_{t-t \wedge \! N} \right|^2 \eta^{t \wedge \! N}
+ \nu \!\! \sum_{i=t-{t \wedge \! N}}^{t-1} \eta^{t-1-i} \left| y_i -
h(\hat x_{i}) \right|^2
+ \omega \!\! \sum_{i=t-{t \wedge \! N}}^{t-1} \eta^{t-1-i} \left| \hat w_{i}
\right|^2 \label{eq:cost:t-N:t-1}
\end{align}
at each time $t=1,2,\ldots$, where $\mu, \nu, \omega > 0$, with implicit constraints
$$
\hat x_{i+1} =  f(\hat x_{i}) + \hat w_{i}  \,,\, i=t\!-\!t \wedge \! N,\ldots,t-1 \,,
$$
and predictions given by \eqref{eq:prediction}. 
Based on the aforesaid, we get the following theorem, which ``mutatis
mutandi'' can be easily derived from 
\cite[Proposition 1 and Theorem 1, p. 7469]{SchiMuntwKohZeilMull23}.

\medskip

\begin{theorem}\label{prop:ISS:perfect:opt}
{\em 
If system \eqref{eq:system} is i-UEIOSS and \eqref{eq:cost:t-N:t-1} is
chosen with $\mu \ge c_x$, $\nu \ge c_v/2$, $\omega \ge c_w$, and 
$N \ge 1$ integer such that $4 \, \mu \, \eta^N < 1$,
the MHE$_N$ resulting from minimizing \eqref{eq:cost:t-N:t-1} is
exponentially robustly stable in that, for each $t=1,2,\ldots$,
there exists $\lambda \in (0,1)$ such that
\begin{align}
&\left| x_t - \hat x_{t|t} \right|^2 \le 4 \, \mu |x_{0} - \bar x_{0}|^2 
\lambda^t + 2 \, \nu \sum_{i=0}^{t-1} \lambda^{t-1-i} |v_i|^2
+ 4 \, \omega \sum_{i=0}^{t-1} \lambda^{t-1-i} |w_i|^2 
\label{eq:robustness:bound:perfect:opt} 
\end{align}
where $\hat x_{t|t}$ is the estimate of $x_t$. 
\qed
}
\end{theorem}

\noindent{\em Proof}. We split the proof into two cases.

First, let us focus on the cases $t=1,\ldots,N-1$. For $t=1$, the minimizer
$\left( \hat x_{0|1} ,
\hat w_{0|1} \right)$ provides the estimate $\hat x_{1|1} = f \left( \hat x_{0|1}
\right) + \hat w_{0|1}$ and hence, by definition, we have $J_1^1 \left( \hat x_{0|1} ,
\hat w_{0|1} \right) \le J_1^1 \left( x_{0} , w_{0} \right)$, i.e.,
\begin{align}
&\mu \left| \hat x_{0|1} - \bar x_{0} \right|^2 \eta + \nu 
\left| y_{0} - h(\hat x_{0|1}) \right|^2 + \omega \left| \hat w_{0|1}
\right|^2 
\le \mu \left| x_{0} - \bar x_{0} \right|^2 \eta + \nu 
\left| y_{0} - h(x_{0}) \right|^2 + \omega \left| w_{0}
\right|^2 \,. \label{eq:lemma:case:1:N-1:first:old}
\end{align}
Using the square sum bound\footnote{The ``square sum bound'' is given by
$$
\left( \sum_{i=1}^m s_i \right)^2 \le m \sum_{i=1}^m s_i^2
$$
for any $s_1, s_2, \ldots , s_m \in \Re$.}, from $\left| x_{0} - \hat x_{0|1} \right|
\le \left| x_{0} - \bar x_{0}  \right| + \left| \bar x_{0} - \hat x_{0|1}
\right|$ it follows that
$$
\left| x_{0} - \hat x_{0|1} \right|^2 \le 2 \left| x_{0} - \bar x_{0} 
\right|^2 + 2 \left| \bar x_{0} - \hat x_{0|1} \right|^2
$$
and hence
\begin{equation}
\frac{1}{2} \left| x_{0} - \hat x_{0|1} \right|^2 - \left| x_{0} - \bar x_{0} 
\right|^2 \le \left| \bar x_{0} - \hat x_{0|1} \right|^2 
\label{eq:ineq:young:x:old}
\end{equation}
and in a similar way
\begin{equation}
\frac{1}{2} \left| w_{0} - \hat w_{0|1} \right|^2 - \left| w_{0} 
\right|^2 \le \left| \hat w_{0|1} \right|^2 \,.
\label{eq:ineq:young:w:old}
\end{equation}
Using \eqref{eq:ineq:young:x:old} and \eqref{eq:ineq:young:w:old}, 
\eqref{eq:lemma:case:1:N-1:first:old} yields
\begin{align}
&\mu \left| x_{0} - \hat x_{0|1} \right|^2 \eta + 2 \, \nu 
\left| y_{0} - h(\hat x_{0|1}) \right|^2 + \omega \left| w_{0}
- \hat w_{0|1} \right|^2
\le 4 \, \mu \left| x_{0} - \bar x_{0} \right|^2 + 2 \, \nu 
\left| v_{0} \right|^2 + 4 \,\omega \left| w_{0} \right|^2 \notag
\end{align}
and hence
\begin{align}
c_x \left| x_{0} - \hat x_{0|1} \right|^2 \eta + c_v
\left| h( x_{0}) + v_{0} - h(\hat x_{0|1}) \right|^2 + c_w
\big| w_{0} 
- \hat w_{0|1} \big|^2 \le 4 \, \mu \left| x_{0} - \bar x_{0} \right|^2 + 2 \,
\nu  \left| v_{0} \right|^2 + 4 \, \omega \left| w_{0} \right|^2 
\label{eq:ineq:almost:last:t=1}
\end{align}
because of the assumptions on $\mu $, $\nu $, and $\omega $.
Thus, by using the i-UEIOSS assumption, from 
\eqref{eq:ineq:almost:last:t=1} it follows that
\begin{align}
&\left| x_1 - \hat x_{1|1} \right|^2 \le 4 \, \mu  \left| x_0 
- \bar x_0 \right|^2 \eta + 2 \, \nu  |v_0|^2 + 4 \, \omega |w_0|^2 \,,
\label{eq:th:1:proof:general:bound:t=1}
\end{align}
namely \eqref{eq:robustness:bound:perfect:opt} for $t=1$ with $\lambda=\eta$.
We can proceed similarly for $t=2,\ldots,N-1$. Thus, we obtain
\begin{align}
&\left| x_t - \hat x_{t|t} \right|^2 \le 4 \, \mu |x_{0} - \bar x_{0}|^2 
\eta^t + 2 \, \nu \sum_{i=0}^{t-1} \eta^{t-1-i} |v_i|^2 
+ 4 \, \omega \sum_{i=0}^{t-1} \eta^{t-1-i} |w_i|^2 \,,\, t=1,\ldots,N-1
\,.
\label{eq:robustness:bound:perfect:opt:t=1:N-1}
\end{align}

Second, from now
on we focus on the cases $t=N,N+1,\ldots$. From the definition of minimum it
follows that 
$J_t^N(\hat x_{t-N|t},\hat w_{t-N|t}^{t-1}) \le J_t^N(x_{t-N},w_{t-N}^{t-1})$, namely
\begin{align}
&\mu \, \left| \hat x_{t-N|t} - \bar x_{t-N} \right|^2 \eta^N + \nu \sum_{i=t-N}^{t-1} 
\eta^{t-1-i} \left| y_i - h(\hat x_{i|t}) \right|^2 
+ \omega \sum_{i=t-N}^{t-1} \eta^{t-1-i} \left| \hat w_{i|t} \right|^2 \le
\mu \, \left| x_{t-N} - \bar x_{t-N} \right|^2 \eta^N \notag \\
&+ \nu \sum_{i=t-N}^{t-1} \eta^{t-1-i} |v_{i}|^2
+ \omega \sum_{i=t-N}^{t-1} \eta^{t-1-i} |w_{i}|^2 
\label{eq:th:1:proof:initial}
\end{align}
for $t=N,N+1,\ldots$. Using the inequalities 
\begin{align}
&\frac{1}{2} \left| x_{t-N} - \hat x_{t-N|t} \right|^2 \! - \left| x_{t-N}
- \bar x_{t-N} \right|^2 \! \le \! \left| \hat x_{t-N|t} - \bar x_{t-N}  \right|^2
\,, \notag \\
&\left|w_{i} - \hat w_{i|t} \right|^2 /2 - \left| w_{i} \right|^2 \le \left|
\hat w_{i|t} \right|^2 \,,\, i=t-N,\ldots,t-1 \,,
\notag 
\end{align}
\eqref{eq:th:1:proof:initial} yields
\begin{align}
&\frac{\mu}{2} \,  \left| x_{t-N} - \hat x_{t-N|t} \right|^2 \eta^N + \nu 
\sum_{i=t-N}^{t-1} \eta^{t-1-i} \left| y_i - h(\hat x_{i|t}) \right|^2
+ \frac{\omega}{2} \sum_{i=t-N}^{t-1} \eta^{t-1-i} \left| w_i - \hat w_{i|t}
\right|^2 \le 2 \mu \, \left| x_{t-N} - \bar x_{t-N} \right|^2 \eta^N \notag \\
&+ \nu \sum_{i=t-N}^{t-1} \eta^{t-1-i} |v_{i}|^2
+ 2 \, \omega \sum_{i=t-N}^{t-1} \eta^{t-1-i} |w_{i}|^2 \,.
\notag 
\end{align}
Since $\mu \ge c_x$, $\nu \ge c_v/2$, $\omega \ge c_w$ and
the system is i-UEIOSS by assumption, we obtain
\begin{align}
&\left| x_{t} - \hat x_{t|t} \right|^2 \le 4 \, \mu  \left| x_{t-N} 
- \bar x_{t-N} \right|^2 \eta^N 
+ 2 \, \nu  \sum_{i=t-N}^{t-1} \eta^{t-1-i} |v_{i}|^2
+ 4 \, \omega \sum_{i=t-N}^{t-1} \eta^{t-1-i} |w_{i}|^2
\label{eq:th:1:proof:general:bound:t>=N}
\end{align}
for all $t=N,N+1,\ldots$. In particular, from \eqref{eq:th:1:proof:general:bound:t>=N}
it follows that
\begin{align}
&\left| x_N - \hat x_{N|N} \right|^2 \le 4 \, \mu  \left| x_0 
- \bar x_0 \right|^2 \eta^N
+ 2 \, \nu  \sum_{i=0}^{N-1} \eta^{N-1-i} |v_{i}|^2
+ 4 \, \omega \sum_{i=0}^{N-1} \eta^{N-1-i} |w_{i}|^2 
\label{eq:th:1:proof:general:bound:t=N}
\end{align}
for $t=N$ and
\begin{align}
&\left| x_{N+1} - \hat x_{N+1|N+1} \right|^2 \le 4 \, \mu  \left| x_1 
- \hat x_{1|1} \right|^2 \eta^N 
+ 2 \, \nu  \sum_{i=1}^{N} \eta^{N-i} |v_{i}|^2
+ 4 \, \omega \sum_{i=1}^{N} \eta^{N-i} |w_{i}|^2
\label{eq:th:1:proof:general:bound:t=N+1}
\end{align}
for $t=N+1$. Using \eqref{eq:th:1:proof:general:bound:t=1}, from
\eqref{eq:th:1:proof:general:bound:t=N+1} we obtain
\begin{align}
\left| x_{N+1} - \hat x_{N+1|N+1} \right|^2 &\le 4 \, \mu \, \left(
4 \, \mu \, \eta^{N+1} \right) \left| x_0 - \bar x_0 \right|^2 
+ 2 \, \nu \left( 4 \, \mu \, \eta^N \left| v_{0} \right|^2 
+ \sum_{i=1}^{N} \eta^{N-i} |v_{i}|^2 \right) \notag \\
&+ 4 \, \omega \left( 4 \, \mu \, \eta^N \left| w_{0} \right|^2 
+ \sum_{i=1}^{N} \eta^{N-i} |w_{i}|^2 \right) \,.
\label{eq:th:1:proof:general:bound:t=N+1:next}
\end{align}
Based on the assumption $4 \, \mu \, \eta^N < 1$, for the theorem of sign
permanence there exists $\alpha \in (0,1)$ such that $4 \, \mu \, 
\eta^{\alpha N} < 1$. Moreover, $\eta < \eta^\alpha < 1$ and $\eta < \eta^{1-\alpha} < 1$. Thus, we get
$$
4 \, \mu \, \eta^N = \underbrace{4 \, \mu \, \eta^{\alpha N}}_{<1} \, 
\eta^{(1-\alpha) N} < \left( \eta^{1-\alpha} \right)^N
$$
and
$$
4 \, \mu \, \eta^{N+1} = \underbrace{4 \, \mu \, \eta^{\alpha (N+1)}}_{<1} \, 
\eta^{(1-\alpha) (N+1)} < \left( \eta^{1-\alpha} \right)^{N+1} \,.
$$
Using such inequalities, from \eqref{eq:th:1:proof:general:bound:t=N+1:next}
it follows that
\begin{align}
&\left| x_{N+1} - \hat x_{N+1|N+1} \right|^2 \le 4 \, \mu \, 
\left| x_0 - \bar x_0 \right|^2 \left( \eta^{1-\alpha} \right)^{N+1}
+ 2 \, \nu \sum_{i=0}^{N} \left( \eta^{1-\alpha} \right)^{N-i} 
|v_{i}|^2 + 4 \, \omega \sum_{i=0}^{N} \left( \eta^{1-\alpha} 
\right)^{N-i} |w_{i}|^2 
\label{eq:th:1:proof:general:bound:t=N+1:final}
\end{align}
and thus \eqref{eq:robustness:bound:perfect:opt} holds for $t=N+1$ with
$\lambda=\eta^{1-\alpha}$. Moreover, it is straightforward to turn
\eqref{eq:robustness:bound:perfect:opt:t=1:N-1} and
\eqref{eq:th:1:proof:general:bound:t=N} into 
\eqref{eq:robustness:bound:perfect:opt} for $t=1,\ldots,N$ with
$\lambda=\eta^{1-\alpha}$ by using the previous bounds. We can go on
in a similar way up to $t=2N-1$. In more detail, let us consider
\eqref{eq:th:1:proof:general:bound:t>=N} for $t=2N$, i.e.,
\begin{align}
&\left| x_{2N} - \hat x_{2N|2N} \right|^2 \le 4 \, \mu  \left| x_N 
- \hat x_{N|N} \right|^2 \eta^N 
+ 2 \, \nu  \sum_{i=N}^{2N-1} \eta^{2N-1-i} |v_{i}|^2
+ 4 \, \omega \sum_{i=N}^{2N-1} \eta^{2N-1-i} |w_{i}|^2 \,,
\label{eq:th:1:proof:general:bound:t=2N}
\end{align}
and, by using \eqref{eq:th:1:proof:general:bound:t=N}, we get
\begin{align}
\left| x_{2N} - \hat x_{2N|2N} \right|^2 &\le 4 \, \mu  \, \left(
4 \, \mu \, \eta^{2N} \right) \left| x_0 
- \bar x_0 \right|^2 
+ 2 \, \nu \left( 4 \, \mu \, \eta^N 
\sum_{i=0}^{N-1} \eta^{N-1-i} |v_{i}|^2
+ \sum_{i=N}^{2N-1} \eta^{2N-1-i} |v_{i}|^2
\right) \notag \\
&+ 4 \, \omega \left( 4 \, \mu \, \eta^N 
\sum_{i=0}^{N-1} \eta^{N-1-i} |w_{i}|^2
\sum_{i=N}^{2N-1} \eta^{2N-1-i} |w_{i}|^2 
\right) \notag
\end{align}
and, following the same treatment bringing to
\eqref{eq:th:1:proof:general:bound:t=N+1:final} from
\eqref{eq:th:1:proof:general:bound:t=N+1:next},
\begin{align}
\left| x_{2N} - \hat x_{2N|2N} \right|^2 &\le 4 \, \mu  
\left| x_0  - \bar x_0 \right|^2 \left( \eta^{1-\alpha} \right)^{2N}
+ 2 \, \nu \sum_{i=0}^{2N-1} \left( \eta^{1-\alpha} \right)^{2N-1-i} |v_{i}|^2 + 4 \, \omega
\sum_{i=0}^{2N-1} \left( \eta^{1-\alpha} \right)^{2N-1-i} |w_{i}|^2 
\,, \notag
\end{align} 
and again \eqref{eq:robustness:bound:perfect:opt} for $t=2N$ with
$\lambda=\eta^{1-\alpha}$. Thus, in general let $t \ge 2N+1$ and suppose that
\begin{align}
&\left| x_{t-N} - \hat x_{t-N|t-N} \right|^2 \le 4 \, \mu  \left| x_0
- \bar x_0 \right|^2 \lambda^{t-N}
+ 2 \, \nu  \sum_{i=0}^{t-N-1} \lambda^{t-N-1-i} |v_{i}|^2
+ 4 \, \omega \sum_{i=0}^{t-N-1} \lambda^{t-N-1-i} |w_{i}|^2
\label{eq:th:1:proof:general:bound:t>=N:from:0}
\end{align}
holds. If we replace \eqref{eq:th:1:proof:general:bound:t>=N:from:0} in
\eqref{eq:th:1:proof:general:bound:t>=N}, we get 
\begin{align}
\left| x_{t} - \hat x_{t|t} \right|^2 &\le 4 \, \mu  \, \left(
4 \, \mu \, \eta^N \right) \left| x_0 
- \bar x_0 \right|^2 \lambda^{t-N} + 2 \, \nu  \left( 4 \, \mu \, \eta^N
\sum_{i=0}^{t-N-1} \lambda^{t-N-1-i} |v_{i}|^2 + \!\!\!
\sum_{i=t-N}^{t-1} \eta^{t-1-i} |v_{i}|^2
\right) \notag \\
&+ 4 \, \omega \left( 4 \, \mu \, \eta^N
\sum_{i=0}^{t-N-1} \lambda^{t-N-1-i} |w_{i}|^2 + \!\!\!
\sum_{i=t-N}^{t-1} \eta^{t-1-i} |w_{i}|^2 
\right) \notag
\end{align}
and, as $4 \, \mu \, \eta^N = \underbrace{4 \, \mu \, \eta^{\alpha N}}_{<1}
\eta^{(1-\alpha) N} < \lambda^N$, 
\begin{align}
&\left| x_{t} - \hat x_{t|t} \right|^2 \le 4 \, \mu \left| x_0 
- \bar x_0 \right|^2 \lambda^t + 2 \, \nu  
\sum_{i=0}^{t-1} \lambda^{t-1-i} |v_{i}|^2 
+ 4 \, \omega \sum_{i=0}^{t-1} \lambda^{t-1-i} |w_{i}|^2 \,,
\notag
\end{align}
thus allowing to conclude.
\qed

\medskip

Results similar to Theorem \ref{prop:ISS:perfect:opt} but under assumptions
of Lipschitz continuity are presented in
\cite[Theorem 3, p. 3481]{KnuMull18} and \cite[Theorem 3, p. 2205]{SchiMull23}. 
The minimization of the cost function can be carried out by means of
some descent method at each time instant under additional assumptions
as follows.

\medskip

\begin{assumption}\label{ass:cont:diff:cost}
{\em The functions $f(\cdot)$ and $h(\cdot)$ 
in \eqref{eq:system}
are continuously differentiable.
}
\end{assumption}

\medskip

\begin{assumption}\label{ass:convex:cost}
{\em The functions $\left(\hat x_0 , \hat w_0^{t-1}
\right) \mapsto
J_t^t(\hat x_0 , \hat w_0^{t-1}) \in \Re_{\ge 0}$ 
for $t=1,2,\ldots,N$ are convex on the convex hull of
$\Bbb X \times \Bbb W^t$.
}
\end{assumption}

\medskip

The optimization required by an MHE$_N$ can be accomplished by using,
for example, the gradient method
with the minimization rule at each time $t$, as shown
in Algorithm \ref{alg:min:rule:grad}, where
$\widetilde x_k \in \Re^n$ and $\widetilde w_{k|t} 
\in \Re^{n t \wedge \! N}$.

\begin{algorithm}
\caption{(gradient method with minimization rule)}
\label{alg:min:rule:grad}
\begin{algorithmic}
\REQUIRE $\bar x_{t-t \wedge \! N}$, $y_{t-t \wedge \! N}^t$, $f(\cdot)$,
$J_t^{t \wedge \! N}(\cdot,\cdot)$, $\nabla J_t^{t \wedge \! N}(\cdot,\cdot)$ 
\STATE $\widetilde x_0 \leftarrow \bar x_{t-t \wedge \! N}$
\STATE $\widetilde w_0 \leftarrow 0$
\STATE $k \leftarrow 0$
\WHILE{
$\left|
\nabla J_t^{t \wedge \! N}\left(\widetilde x_k , \widetilde w_k \right)
\right| \not= 0$
}
\STATE $\alpha_k \leftarrow \arg \dsst \min_{\alpha \ge 0} \left\{
J_t^{t \wedge \! N}\left( \left(\widetilde x_k , \widetilde w_k \right) - 
\alpha \, \nabla J_t^{t \wedge \! N}\left(\widetilde x_k , \widetilde w_k \right)
\right) \right\}$
\STATE $\left(\widetilde x_{k+1} , \widetilde w_{k+1} \right) \leftarrow
\left(\widetilde x_k , \widetilde w_k \right) - \alpha_k \, 
\nabla J_t^{t \wedge \! N}\left(\widetilde x_k , \widetilde w_k \right)$
\STATE $k \leftarrow k+1$
\ENDWHILE
\STATE $\hat x_{t-t \wedge \! N|t} \leftarrow \widetilde x_k$
\STATE $\hat w_{t-t \wedge \! N|t}^{t-1} \leftarrow \widetilde w_k$
\FOR{$i=0$ to $t \wedge \! N-1$}
\STATE $\hat x_{t-t \wedge \! N+i+1|t} \leftarrow f\left(
\hat x_{t-t \wedge \! N+i|t}\right) + \hat w_{t-t \wedge \! N+i|t}$
\ENDFOR
\ENSURE $\hat x_{t|t}$
\end{algorithmic}
\end{algorithm}

In Algorithm \ref{alg:min:rule:grad} we get 
$\left( \hat x_{t-t \wedge \! N|t} \,,\,
\hat w_{t-t \wedge \! N|t}^{t-1} \right)$ as a stationary point of the
minimization problem, i.e.,
\begin{equation}
\nabla J_t^{t \wedge \! N} \left( \hat x_{t-t \wedge \! N|t} \,,\,
\hat w_{t-t \wedge \! N|t}^{t-1} \right) = 0 \,,\,
t=1,2,\ldots \,. \label{eq:grad:null:cost}
\end{equation}
Because of Assumptions \ref{ass:cont:diff:cost} and \ref{ass:convex:cost} 
it follows that
\begin{align}
&J_t^{t \wedge \! N} \!\! \left( x_{t-t \wedge \! N|t} \,,\, w_{t-t \wedge \! N|t}^{t-1}
\right) \ge J_t^{t \wedge \! N} \!\! \left( \hat x_{t-t \wedge \! N|t} \,,\,
\hat w_{t-t \wedge \! N|t}^{t-1} \right) 
+  \nabla J_t^{t \wedge \! N} \!\! \left( \hat x_{t-t \wedge \! N|t} \,,\,
\hat w_{t-t \wedge \! N|t}^{t-1} \right)^{\!\!\top} \!\!
\left( \!
\begin{array}{c}
x_{t-t \wedge \! N|t} - \hat x_{t-t \wedge \! N|t} \\
\hdashline\\[-0.3cm]
w_{t-t \wedge \! N|t}^{t-1} - \hat w_{t-t \wedge \! N|t}^{t-1}
\end{array}
\! \right) \label{eq:grad:cost:convex:bound}
\end{align} 
and, by using \eqref{eq:grad:null:cost},
$$
J_t^{t \wedge \! N} \!\! \left( \hat x_{t-t \wedge \! N|t} \,,\, \hat w_{t-t 
\wedge N|t}^{t-1} \right) \le J_t^{t \wedge \! N} \!\! \left( x_{t-t \wedge \! N|t}
\,,\, w_{t-t \wedge \! N|t}^{t-1} \right)
$$
and thus the robust exponential stability of the estimation error
is proved by arguing like in the proof of
Theorem \ref{prop:ISS:perfect:opt}.

\medskip

\begin{remark}
Convexity is explicitly required in this paper for what follows in Section \ref{sec:imperfect:opt}.
In practice, this assumption holds because of observability when dealing with moving-horizon estimation
for linear systems \cite[p. 4505]{AleGagTAC17} since observability ensures that the
quadratic cost function is positive definite and thus convex.
Since, as pretty well known, the composition of convex functions is convex, we are not limited to
a linear state equation with a quadratic cost function.
For nonlinear systems, convexity
can be proved to hold by checking that the Hessian matrix of the cost function is
positive definite. If the cost function is not smooth enough, one can use a method
based on mathematical programming. Specifically, consider
$g : D \rightarrow \Re$ with  $D \subset \Re^n$ convex. If the mathematical programming
problem
\begin{align}
&\max \left| \xi-\zeta\right| \quad \mbox{w.r.t.} \quad \xi,\zeta \in D, \tau \in (0,1) \quad \mbox{s.t.} \quad \notag \\
&\tau \, g(\xi) + (1-\tau) \, g(\zeta) - g\left( \tau \xi + (1-\tau) \zeta \right) \le 0 \notag
\end{align}
provides $\xi^*, \zeta^*, \tau^*$ as solution with  $\xi^* \not= \zeta^*$, 
then $g(\cdot)$ is not convex; by contrast,
if $\xi^* \equiv \zeta^*$, $g(\cdot)$ is convex. Clearly, it is a feasibility program that aims
at satisfying an inequality condition, which certifies that convexity does not hold.
\end{remark}

\medskip

In the next section, we will consider to perform moving-horizon estimation
under a more relaxed stopping criterion as compared with the null
gradient.

\section{Moving-horizon Estimation with Imperfect Optimization}
\label{sec:imperfect:opt}

We leverage Algorithm \ref{alg:min:rule:grad:imperfect}, where 
we still refer to the gradient method with minimization rule like
in Algorithm \ref{alg:min:rule:grad} but exiting the loop with
\begin{equation}
\left|
\nabla J_t^{t \wedge \! N} \! \left( \widetilde x_k , \widetilde w_{k|t}
\right) \right| \le \hat \varepsilon_t \label{eq:TC:norm:grad:small}
\end{equation}
as {\em termination criterion} for some $\hat \varepsilon_t > 0$ to be suitably chosen,
instead of demanding a null gradient as in Algorithm \ref{alg:min:rule:grad}.
In other words, the search of the optimum is stopped if the norm of the gradient is
less than $\hat \varepsilon_t$, thus reducing the number of iterations.

\medskip

\begin{algorithm}
\caption{(gradient method with minimization rule and relaxed stopping
criterion)}
\label{alg:min:rule:grad:imperfect}
\begin{algorithmic}
\REQUIRE $\bar x_{t-t \wedge \! N}$, $y_{t-t \wedge \! N}^t$, $f(\cdot)$,
$J_t^{t \wedge \! N}(\cdot,\cdot)$, $\nabla J_t^{t \wedge \! N}(\cdot,\cdot)$, $\hat \varepsilon_t$
\STATE $\widetilde x_0 \leftarrow \bar x_{t-t \wedge \! N}$
\STATE $\widetilde w_0 \leftarrow 0$
\STATE $k \leftarrow 0$
\WHILE{
$\left|
\nabla J_t^{t \wedge \! N}\left(\widetilde x_k , \widetilde w_k \right)
\right| > \hat \varepsilon_t$
}
\STATE $\alpha_k \leftarrow \arg \dsst \min_{\alpha \ge 0} \left\{
J_t^{t \wedge \! N}\left( \left(\widetilde x_k , \widetilde w_k \right) - 
\alpha \, \nabla J_t^{t \wedge \! N}\left(\widetilde x_k , \widetilde w_k \right)
\right) \right\}$
\STATE $\left(\widetilde x_{k+1} , \widetilde w_{k+1} \right) \leftarrow
\left(\widetilde x_k , \widetilde w_k \right) - \alpha_k \, 
\nabla J_t^{t \wedge \! N}\left(\widetilde x_k , \widetilde w_k \right)$
\STATE $k \leftarrow k+1$
\ENDWHILE
\STATE $\hat x_{t-t \wedge \! N|t} \leftarrow \widetilde x_k$
\STATE $\hat w_{t-t \wedge \! N|t}^{t-1} \leftarrow \widetilde w_k$
\FOR{$i=0$ to $t \wedge \! N-1$}
\STATE $\hat x_{t-t \wedge \! N+i+1|t} \leftarrow f\left(
\hat x_{t-t \wedge \! N+i|t}\right) + \hat w_{t-t \wedge \! N+i|t}$
\ENDFOR
\ENSURE $\hat x_{t|t}$
\end{algorithmic}
\end{algorithm}

\medskip

\begin{theorem}\label{theo:ISS:imperfect:opt}
{\em 
If system \eqref{eq:system} is i-UEIOSS and Assumptions 
\ref{ass:cont:diff:cost}-\ref{ass:convex:cost} hold, the MHE$_N$ based on 
\eqref{eq:cost:t-N:t-1}
with $N \ge 1$ integer and parameters $\mu, \nu$, and $\omega$ such that
\begin{align}
&\mu \ge c_x + 1 \;,\; \nu \ge
\frac{c_v}{2} \;,\; \omega \ge c_w + \eta
\label{eq:cost:imperf:opt:parameters:cond:mu:nu:omega:lower:bound} \\
&4 \, \mu \, \eta^N < 1
\label{eq:cost:imperf:opt:parameters:cond:contraction}
\end{align}
and by using  Algorithm \ref{alg:min:rule:grad:imperfect} with 
\begin{equation}
\hat \varepsilon_t = \left\{
\begin{array}{ll}
\varepsilon \, \eta^{t/2}        & t=1,\ldots,N \\
\varepsilon \, \eta^{N/2} \sqrt{1-4 \, \mu \, \eta^N} & t=N+1,N+2,\ldots
\end{array}
\right.
\label{eq:hat:epsilon:t}
\end{equation}
for some $\varepsilon > 0$ is $\varepsilon$-practically exponentially
robustly stable in that, for each $t=1,2,\ldots$, there exists $\lambda \in (0,1)$ such that
\begin{align}
&\left| x_t - \hat x_{t|t} \right|^2 \le 
4 \, \mu  |x_{0} - \bar x_{0}|^2 \lambda^t + 2
\nu \sum_{i=0}^{t-1} \lambda^{t-1-i} |v_i|^2 
+ 4 \, \omega \sum_{i=0}^{t-1} \lambda^{t-1-i} |w_i|^2
+ \varepsilon^2 \label{eq:robustness:bound:imperf:opt}
\end{align}
where $\hat x_{t|t}$ is the estimate of $x_t$.
}
\end{theorem}

\noindent{\em Proof}. Let us consider \eqref{eq:grad:cost:convex:bound}
and Algorithm \ref{alg:min:rule:grad:imperfect}, from which we get
\begin{align}
J_t^{t \wedge \! N} \!\! &\left( \hat x_{t-t \wedge \! N|t} \,,\,
\hat w_{t-t \wedge \! N|t}^{t-1} \right) \le J_t^{t \wedge \! N} \!\! 
\left( x_{t-t \wedge \! N|t} \,,\, w_{t-t \wedge \! N|t}^{t-1}
\right) 
+ \nabla J_t^{t \wedge \! N} \!\! \left( \hat x_{t-t \wedge \! N|t} \,,\,
\hat w_{t-t \wedge \! N|t}^{t-1} \right)^{\!\!\top} \!\!
\left( \!
\begin{array}{c}
\hat x_{t-t \wedge \! N|t} - x_{t-t \wedge \! N|t} \\
\hdashline
\hat w_{t-t \wedge \! N|t}^{t-1} - w_{t-t \wedge \! N|t}^{t-1}
\end{array}
\! \right) \notag \\
&\le J_t^{t \wedge \! N} \!\! 
\left( x_{t-t \wedge \! N|t} \,,\, w_{t-t \wedge \! N|t}^{t-1}
\right) + \hat \varepsilon_t \Big| \Big( \hat x_{t-t \wedge \! N|t} -
x_{t-t \wedge \! N|t} , \hat w_{t-t \wedge \! N|t}^{t-1} - 
w_{t-t \wedge \! N|t}^{t-1} \Big)
\Big| \,,\, t=1,2,\ldots
\label{eq:grad:cost:convex:bound:revised}
\end{align}
by means of the Schwarz's inequality and \eqref{eq:TC:norm:grad:small}.
Using the Young's inequality,
\eqref{eq:grad:cost:convex:bound:revised} yields
\begin{align}
J_t^{t \wedge \! N} \!\! \left( \hat x_{t-t \wedge \! N|t} \,,\,
\hat w_{t-t \wedge \! N|t}^{t-1} \right) &\le J_t^{t \wedge \! N} \!\! 
\left( x_{t-t \wedge \! N|t} \,,\, w_{t-t \wedge \! N|t}^{t-1}
\right)
+ \frac{\hat \varepsilon_t^2}{2 \, \eta^{t \wedge \! N}} + 
\frac{\eta^{t \wedge \! N}}{2}
\bigg( \left| x_{t-t \wedge \! N|t} - \hat x_{t-t \wedge \! N|t} \right|^2 +
\Big| w_{t-t \wedge \! N|t}^{t-1} \notag \\
&- \hat w_{t-t \wedge \! N|t}^{t-1} \Big|^2 \bigg) \,,\, t=1,2,\ldots
\,.
\label{eq:grad:cost:bound:after:Schwarz:Young}
\end{align}
From \eqref{eq:grad:cost:bound:after:Schwarz:Young}
for $t=1$, it follows that
$$
J_1^1 \left( \hat x_{0|1} , \hat w_{0|1} \right) \notag \le J_1^1 \left( x_{0} ,
w_{0} \right) + \frac{\varepsilon_1^2}{2 \, \eta} + \frac{\eta}{2}
\bigg( \left| x_0 - \hat x_{0|1} \right|^2 
+ \left| w_0  - \hat w_{0|1} \right|^2 \bigg) 
$$
with $\hat x_{1|1} = f \left( \hat x_{0|1}
\right) + \hat w_{0|1}$ and $\varepsilon_1 = \varepsilon \, \sqrt{\eta}$,
and thus
\begin{align}
\mu \left| \hat x_{0|1} - \bar x_{0} \right|^2 \eta + \nu 
\left| y_{0} - h(\hat x_{0|1}) \right|^2 + \omega \left| \hat w_{0|1}
\right|^2 
&\le \mu \left| x_{0} - \bar x_{0} \right|^2 \eta + \nu 
\left| v_{0} \right|^2 + \omega \left| w_{0}
\right|^2 
+ \frac{\varepsilon^2}{2} \notag \\
&+ \frac{\eta}{2} \bigg( \left| x_0 - \hat x_{0|1} \right|^2 +
\left| w_0  - \hat w_{0|1} \right|^2 \bigg)
\,. \label{eq:lemma:case:1:N-1:first}
\end{align}
Using the square sum bound, from $\left| x_{0} - \hat x_{0|1} \right|
\le \left| x_{0} - \bar x_{0}  \right| + \left| \bar x_{0} - \hat x_{0|1}
\right|$ it follows that
$$
\left| x_{0} - \hat x_{0|1} \right|^2 \le 2 \left| x_{0} - \bar x_{0} 
\right|^2 + 2 \left| \bar x_{0} - \hat x_{0|1} \right|^2
$$
and hence
\begin{equation}
\frac{1}{2} \left| x_{0} - \hat x_{0|1} \right|^2 - \left| x_{0} - \bar x_{0} 
\right|^2 \le \left| \hat x_{0|1} - \bar x_{0} \right|^2 
\label{eq:ineq:young:x}
\end{equation}
and in a similar way
\begin{equation}
\frac{1}{2} \left| w_{0} - \hat w_{0|1} \right|^2 - \left| w_{0} 
\right|^2 \le \left| \hat w_{0|1} \right|^2 \,.
\label{eq:ineq:young:w}
\end{equation}
Using \eqref{eq:ineq:young:x} and \eqref{eq:ineq:young:w}
in \eqref{eq:lemma:case:1:N-1:first}, we get
\begin{align}
\left( \frac{\mu}{2} - \frac{1}{2} \right) 
\left| x_{0} - \hat x_{0|1} \right|^2 \eta + \nu 
\left| y_{0} - h(\hat x_{0|1}) \right|^2
+ \left( \frac{\omega}{2} -
\frac{\eta}{2} \right) \left| w_0 - \hat w_{0|1} \right|^2 
&\le 2 \, \mu \left| x_{0} - \bar x_{0} \right|^2 \eta \notag \\
&+ \nu \left| v_{0} \right|^2 + 2 \, \omega \left| w_{0}
\right|^2 + \frac{\varepsilon^2}{2}
\notag
\end{align}
and finally
\begin{align}
\left( \mu - 1 \right) \left| x_{0} - \hat x_{0|1}
\right|^2 \eta + 2 \, \nu  \left| y_{0} - h(\hat x_{0|1}) \right|^2
+ \left( \omega - \eta \right) \left| w_0 - \hat w_{0|1} \right|^2
&\le 4 \, \mu \left| x_{0} - \bar x_{0} \right|^2 \eta + 2 \, \nu 
\left| v_{0} \right|^2 + 4 \, \omega \left| w_{0}
\right|^2 \notag \\
&+ \varepsilon^2 \,. 
\label{eq:grad:cost:bound:after:t:1:next}
\end{align}
Since system \eqref{eq:system} is i-UEIOSS, by using
\eqref{eq:cost:imperf:opt:parameters:cond:mu:nu:omega:lower:bound} from
\eqref{eq:grad:cost:bound:after:t:1:next} we obtain
\begin{align}
\left| x_{1} - \hat x_{1|1} \right|^2 \! \le \!
4 \, \mu \left| x_{0} - \bar x_{0} \right|^2 \eta + 2 \, \nu 
\left| v_{0} \right|^2 + 4 \, \omega \left| w_{0} \right|^2 \! + \!
\varepsilon^2 
\label{eq:grad:cost:bound:after:t:1:final}
\end{align}
and thus \eqref{eq:robustness:bound:imperf:opt} holds for $t=1$ with $\lambda=\eta$.
It is straightforward to proceed for $t=2,\ldots,N$ in a similar way.
In particular, for $t=N$ from
\begin{align}
&J_N^N \left( \hat x_{0|N} , \hat w_{0|N}^{N-1} \right) \le J_N^N \left( x_0 ,
w_0^{N-1} \right) + \frac{\varepsilon^2}{2 \, \eta^N} + \frac{\eta^N}{2}
\bigg( \left| x_0 - \hat x_{0|N} \right|^2 
+ \left| w_0^{N-1}  - \hat w_{0|N}^{N-1} \right|^2 \bigg) 
\notag
\end{align}
and, since 
\begin{align}
\left| w_i - \hat w_{i|N} \right|^2 &= \frac{\eta^{N-1-i}}{\eta^{N-1-i}}
\left| w_i - \hat w_{i|N} \right|^2 
\le \frac{\eta^{N-1-i}}{\eta^{N-1}} \left| w_i - \hat w_{i|N} \right|^2
\,,\, i=0,\ldots,N-1
\notag
\end{align}
and thus
$$
\eta^N \left| w_0^{N-1}  - \hat w_{0|N}^{N-1} \right|^2 \le \eta
\sum_{i=0}^{N-1} \eta^{N-1-i} \left| w_i - \hat w_{i|N} \right|^2
\,,
$$
it follows that
\begin{align}
&\left( \mu - 1 \right) 
\left| x_{0} - \hat x_{0|N} \right|^2 \eta^N 
+ 2 \, \nu  \sum_{i=0}^{N-1} \eta^{N-1-i} \left| y_{i}
- h(\hat x_{i|N}) \right|^2 
+ \left( \omega - \eta \right) \sum_{i=0}^{N-1} \eta^{N-1-i} 
\left| w_i - \hat w_{i|N} \right|^2 \notag \\
&\le 4 \, \mu \left| x_{0} - \bar x_{0} \right|^2 \eta^N
+ 2 \, \nu  \sum_{i=0}^{N-1} \eta^{N-1-i} \left| v_{i} \right|^2
+ 4 \, \omega \sum_{i=0}^{N-1} \eta^{N-1-i} \left| w_{i}
\right|^2 + \varepsilon^2 \,. 
\label{eq:grad:cost:bound:after:t:N-1:next}
\end{align}
Again, since system \eqref{eq:system} is i-UEIOSS and \eqref{eq:cost:imperf:opt:parameters:cond:mu:nu:omega:lower:bound} holds, from
\eqref{eq:grad:cost:bound:after:t:N-1:next} we obtain
\begin{align}
&\left| x_N - \hat x_{N|N} \right|^2
\le 4 \, \mu \left| x_{0} - \bar x_{0} \right|^2 \eta^N
+ 2 \, \nu  \sum_{i=0}^{N-1} \eta^{N-1-i} \left| v_{i} \right|^2
+ 4 \, \omega \sum_{i=0}^{N-1} \eta^{N-1-i} \left| w_{i}
\right|^2 + \varepsilon^2 \,. 
\label{eq:grad:cost:bound:after:t:N-1:final}
\end{align}

Let us now consider $t=N+1,N+2,\ldots$ and, moving from 
\eqref{eq:grad:cost:bound:after:Schwarz:Young} while following
the same previous passages, we get
\begin{align}
&\left( \frac{\mu}{2} - \frac{1}{2} \right) \, 
\left| x_{t-N} - \hat x_{t-N|t} \right|^2 \eta^N
+ \nu \sum_{i=t-N}^{t-1} \eta^{t-1-i} \left| y_i \right.
\left. - h(\hat x_{i|t}) \right|^2 
+ \frac{\omega}{2} 
\sum_{i=t-N}^{t-1} \eta^{t-1-i} \left| w_i - \hat w_{i|t} \right|^2
\notag \\ 
&- \frac{\eta^N}{2} \sum_{i=t-N}^{t-1} \left| w_i - \hat w_{i|t} \right|^2
\le 2 \, \mu \left| x_{t-N} - \bar x_{t-N} \right|^2 \eta^N 
+ \nu \sum_{i=t-N}^{t-1} \eta^{t-1-i} |v_{i}|^2
+ 2 \, \omega \sum_{i=t-N}^{t-1} \eta^{t-1-i} |w_{i}|^2 
+ \frac{\hat \varepsilon_t^2}{2 \, \eta^N} \,.
\label{eq:theo:2:intermediate:bound}
\end{align}
Since
\begin{align}
&\left| w_i - \hat w_{i|t} \right|^2 = \frac{\eta^{t-1-i}}{\eta^{t-1-i}}
\left| w_i - \hat w_{i|t} \right|^2 
\le \frac{\eta^{t-1-i}}{\eta^{N-1}} \left| w_i - \hat w_{i|t} \right|^2 
\,,\, i=t-N,\ldots,t-1 \,, \notag
\end{align}
it follows that
\begin{align}
-\frac{\eta}{2} \sum_{i=t-N}^{t-1} \eta^{t-1-i}
\left| w_i - \hat w_{i|t} \right|^2 \le -\frac{\eta^N}{2} \sum_{i=t-N}^{t-1} \left| w_i - \hat w_{i|t} \right|^2
\,. \label{eq:theo:2:bound:wi-hatwi}
\end{align}
After replacing \eqref{eq:theo:2:bound:wi-hatwi} in 
\eqref{eq:theo:2:intermediate:bound}, we obtain
\begin{align}
&\left( \mu - 1 \right) \, 
\left| x_{t-N} - \hat x_{t-N|t} \right|^2 \eta^N 
+ 2 \, \nu \sum_{i=t-N}^{t-1} \eta^{t-1-i} \left| y_i \right.
\left. - h(\hat x_{i|t})
\right|^2  
+ \left( \omega - \eta
\right) \sum_{i=t-N}^{t-1} \eta^{t-1-i} \left| w_i - \hat w_{i|t} \right|^2
\notag \\
&\le 4 \, \mu \left| x_{t-N} - \bar x_{t-N} \right|^2 \eta^N 
+ 2 \, \nu \sum_{i=t-N}^{t-1} \eta^{t-1-i} |v_{i}|^2
+ 4 \, \omega \sum_{i=t-N}^{t-1} \eta^{t-1-i} |w_{i}|^2 
+ \frac{\hat \varepsilon_t^2}{\eta^N} \notag
\end{align}
and, again by using  \eqref{eq:cost:imperf:opt:parameters:cond:mu:nu:omega:lower:bound}
and since system \eqref{eq:system} is i-UEIOSS,
we get
\begin{align}
\left| x_{t} - \hat x_{t|t} \right|^2 &\le 
4 \, \mu \left| x_{t-N} - \bar x_{t-N} \right|^2 \eta^N 
+ 2 \, \nu \sum_{i=t-N}^{t-1} \eta^{t-1-i} |v_{i}|^2 \notag \\
&+ 4 \, \omega \sum_{i=t-N}^{t-1} \eta^{t-1-i}
|w_{i}|^2 + \frac{\hat \varepsilon_t^2}{\eta^N} \,,\,
t=N+1,N+2,\ldots \,. \label{eq:theo:2:general:bound}
\end{align}
From \eqref{eq:theo:2:general:bound} for $t=N+1$ it follows that
\begin{align}
&\left| x_{N+1} - \hat x_{N+1|N+1} \right|^2 \le 
4 \, \mu \left| x_1 - \hat x_{1|1} \right|^2 \eta^N 
+ 2 \, \nu \sum_{i=1}^{N} \eta^{N-i} |v_{i}|^2 
+ 4 \, \omega \sum_{i=1}^{N} \eta^{N-i}
|w_{i}|^2 + \frac{\hat \varepsilon_{N+1}^2}{\eta^N} \,.
\label{eq:theo:2:bound:t=N+1}
\end{align}
Since, by assumption \eqref{eq:cost:imperf:opt:parameters:cond:contraction}, i.e.,
$4 \, \mu \, \eta^N < 1$, for the theorem of sign
permanence there exists $\alpha \in (0,1)$ such that $4 \, \mu \, 
\eta^{\alpha N} < 1$, we obtain
$$
4 \, \mu \, \eta^{N+1} = \underbrace{4 \, \mu \, \eta^{\alpha (N+1)}}_{<1} \, 
\lambda^{N+1} < \lambda^{N+1}
$$
with $\lambda=\eta^{1-\alpha}$ and hence, by using \eqref{eq:grad:cost:bound:after:t:1:final},
\begin{align}
& \left| x_{N+1} - \hat x_{N+1|N+1} \right|^2 \le 
4 \, \mu \left| x_0 - \bar x_0 \right|^2 \lambda^{N+1}
+ 2 \, \nu \sum_{i=0}^{N} \lambda^{N-i} |v_{i}|^2 
+ 4 \, \omega \sum_{i=0}^{N} \lambda^{N-i}
|w_{i}|^2 + 4 \, \mu \, \eta^N \varepsilon^2 
+ \frac{\hat \varepsilon_{N+1}^2}{\eta^N}\,.
\notag
\end{align}
Thus, we get \eqref{eq:robustness:bound:imperf:opt}
for $t = N+1$ as
$$
\hat \varepsilon_{N+1} = \varepsilon \, \eta^{N/2}
\sqrt{1-4 \, \mu \, \eta^N}
$$
and so on for all $t \ge N+2$ from \eqref{eq:theo:2:general:bound}.
\qed


The core of the proof of Theorem 3 is the selection of cost weights, 
discount factors, parameters in the stopping criterion,
and sufficiently large batch of input-output pairs to fit,
while deriving the appropriate upper bounding.


%

\section{Numerical Results}
\label{sec:sim}

Two case studies are addressed in the following: specifically, we will consider  an i-UEIOSS system
in Section \ref{sec:example:1}, while the system in Section \ref{sec:example:2} is i-UIOSS but not i-UEIOSS.
The solution of all the optimization problems were obtained by using routines
of the Matlab optimization toolbox.

\subsection{i-UEIOSS Example}
\label{sec:example:1}

A system described by
$$
\left\{
\begin{array}{l}
x_{t+1} =  f(x_t,u_t,w_t)  \\
y_t = h(x_t)
\end{array} \,,\, t=0,1,\ldots
\right.
$$
with $x_t \in \Bbb X \subseteq \Re^n$, $u_t \in \Bbb U \subseteq \Re^p$, $y_t \in \Re^m$, 
and $w_t \in \Bbb W \subseteq \Re^n$ is i-UIOSS if and only if there exists
an i-UIOSS Lyapunov function, i.e., some $V : \Bbb X \times \Bbb X \rightarrow \Re_{\ge 0}$ such
that there exist $\alpha_1, \alpha_2, \alpha_3 \in \cK_\infty$,  
$\sigma_w, \sigma_y \in \cK$ satisfying the inequalities
\begin{align}
&\alpha_1(|x-z|) \le V(x,z) \le \alpha_2(|x-z|) \label{eq:example:V:sandwich} \\
&V \left( f(x,u,w_x) , f(z,u,w_z) \right) \le V(x,z) - \alpha_3(|x-z|) 
+ \sigma_w(|w_x-w_z|) + \sigma_y \left( |h(x)-h(z)| \right)
\label{eq:example:V:dissip}
\end{align}
for all $x,z \in \Bbb X$, $u \in \Bbb U$, and $w_x, w_z \in \Bbb W$
(see \cite[Proposition 5, p. 4499]{AllanRawlACC19} and
\cite[Theorem 3.2, p. 3025]{AllanRawlTeel21}, while
using \cite[Theorem B.15, p. 703]{RawlMayneDiehlBOOK17}). If $\alpha_1(s),
\alpha_2(s), \alpha_3(s), \sigma_w(s), \sigma_y(s)$ are quadratic in $s$,
i-UEIOSS holds.

Consider the third-order system with
\begin{equation}\label{eq:example:1}
\begin{array}{l}
f(x,u,w) = \bigg( \dsst
\frac{x_1}{4} + \frac{\log \left( |x_2| + 1 \right)}{4} + w_1 \,,\,
\arctan \left( x_1 + x_3^2 \right) + w_2 \,,\, \dsst
\frac{\sin ( x_2 + x_3 )}{4} + u + w_3 \bigg) \in \Re^3 \\
h(x) = x_1 + x_3^2 \in \Re
\end{array}
\end{equation}
where $u \in \Re$ and $x, w \in \Re^3$ with $x_i, w_i$ denoting the $i$-th component of
$x, w$, respectively. 
Since it is easy to prove for \eqref{eq:example:1} with
$(x,z) \mapsto V(x,z) = (x-z)^\top (x-z)$ that the inequality
\begin{align}
V\left( f(x,u,w_x) , f(z,u,w_z) \right) &\le \frac{7}{16} V(x,z) + 3 |w_x-w_z|^2
+2 (h(x)-h(z))^2 \label{eq:i-UEIOSS:bound:V}
\end{align}
holds for all $x, z, w_x, w_z \in \Re^3$, it follows that $V(\cdot,\cdot)$ is an i-UIOSS
Lyapunov function with $\alpha_1(s)=\alpha_2(s)=s^2$, $\alpha_3(s)= 9/16 s^2$,
$\sigma_w(s) = 3 s^2$, $\sigma_y(s) = 2 s^2$ in \eqref{eq:example:V:sandwich} 
and \eqref{eq:example:V:dissip} and thus this system is i-UEIOSS. 

%
%
Concerning Theorem \ref{theo:RS:general}, since
$$
\sigma(s) = s - \frac{1}{2} \alpha_3 \left( \alpha_2^{-1}(s) \right) =
\lambda s < s
$$
with $\lambda=23/32$, from the proof of \cite[Proposition 5, p. 4499]{AllanRawlACC19}
we obtain 
\begin{align}
&|x_{t} - z_{t}| \le \beta_x(|x_{0}-z_{0}|,t) 
\oplus \Oplus_{i=0}^{t-1} \beta_y ( |h(x_{i})-h(z_{i})| , t\!-\!1\!-\!i )
\oplus \Oplus_{i=0}^{t-1} \beta_w ( |w_{x_i}-w_{z_i}| , t\!-\!1\!-\!i ) \,,\,
t=1,2,\ldots \notag
\end{align}
where
\begin{align}
\beta_x(s,t) &:= \alpha_1^{-1} \! \left( \sigma^t ( \alpha_2 ( s ) ) \right)
= s \, \eta^t \notag \\
\beta_y(s,t) &:= \alpha_1^{-1} \! \left( \sigma^t ( 
2 \alpha_2 ( 
\alpha_3^{-1}(
4 \sigma_w(s)
)
) + 2 \sigma_y(s)
) \right) = \sqrt{140/3} \, s \, \eta^t \notag \\
\beta_w(s,t) &:= \alpha_1^{-1} \! \left( \sigma^t ( 
2 \alpha_2 ( 
\alpha_3^{-1}(
4 \sigma_w(s)
)
) + 2 \sigma_w(s)
) \right) = \sqrt{146/3} \, s \, \eta^t \notag
\end{align}
with
$\sigma^k(s) = \overbrace{\sigma(s) \circ \cdots \circ \sigma(s)}^{k \; 
\mbox{\small times}} = s \, \eta^k $, $k \in \Na_{\ge 1}$,
$\eta:=\sqrt{\lambda}$. Let us verify how \eqref{eq:cond:theo:RS:general:alpha:beta:x} can be
satisfied if $N$ is chosen such that $2 \eta^N < 1$, namely for values of $N$
not less than 5. Toward this end, we have
\begin{align}
&\beta_x \left( 4 \beta_x(2s,k) , N \right) = 8 \, s \, \eta^{k+N}
\label{eq:cond:prop:RS:general:alpha:beta:1:x:example} \\
&\beta_x \left( 2 \alpha_x(s,k) , N \right) = 2 \, \alpha_x(s,k) \, \eta^N \,.
\label{eq:cond:prop:RS:general:alpha:beta:2:x:example} 
\end{align}
If we choose $\alpha_x(s,k) = 8 \, s \, (\eta^c)^k$ with $c \in (0,1)$, it
follows that $\alpha_x(s,k) \ge 8 \, s \, \eta^k$ and hence 
\eqref{eq:cond:theo:RS:general:alpha:beta:1:x} is satisfied. If $2 \eta^N < 1$,
there exists $c \in (0,1)$ such that $2 (\eta^{1-c})^N < 1$ and thus
from \eqref{eq:cond:prop:RS:general:alpha:beta:2:x:example} for this
specific choice of $c$ we get
\begin{align}
&\beta_x \left( 2 \alpha_x(s,k) , N \right) = 16 \, (\eta^{c})^k \eta^N
= 16 \, (\eta^{c})^k (\eta^{c})^N (\eta^{1-c})^N \notag \\
&= 2 (\eta^{1-c})^N \alpha_x(s,k+N) \le \alpha_x(s,k+N) \notag
\end{align}
and so \eqref{eq:cond:theo:RS:general:alpha:beta:2:x} holds. By following
the same reasoning it is easy to show that \eqref{eq:cond:theo:RS:general:alpha:beta:w} and
\eqref{eq:cond:theo:RS:general:alpha:beta:v} are satisfied if $2 \eta^N < 1$.
For the purpose of comparison, note that the robust stability condition of \cite[Theorem 14, p. 6]{KnuMull23}
requires the satisfaction of the same condition, thus demanding a window of length not lower than 5.  
The results of noise-free and noisy simulation runs are shown in Fig. 1
(MHE$_5$ with {\em fminimax}).

%
%
Concerning Theorem \ref{theo:ISS:imperfect:opt}, from \eqref{eq:i-UEIOSS:bound:V} it is straightforward
to show that \eqref{eq:def:i-UEIOSS:bound} holds with $c_x=1$, $c_v = 140/3$, $c_w = 146/3$, and
$\eta=7/16$. Thus, \eqref{eq:cost:imperf:opt:parameters:cond:mu:nu:omega:lower:bound}
and \eqref{eq:cost:imperf:opt:parameters:cond:contraction} are satisfied by
$\mu=2, \nu=70/3, \omega=c_w+\eta=2357/48$, and any integer $N \ge 3$. 
Simulation results are presented in Fig. 1 (MHE$_3$ with {\em fmincon}). 

\begin{figure*}[htb]
\label{fig:simul:example:1}
\centering
\includegraphics[width=16.5cm]{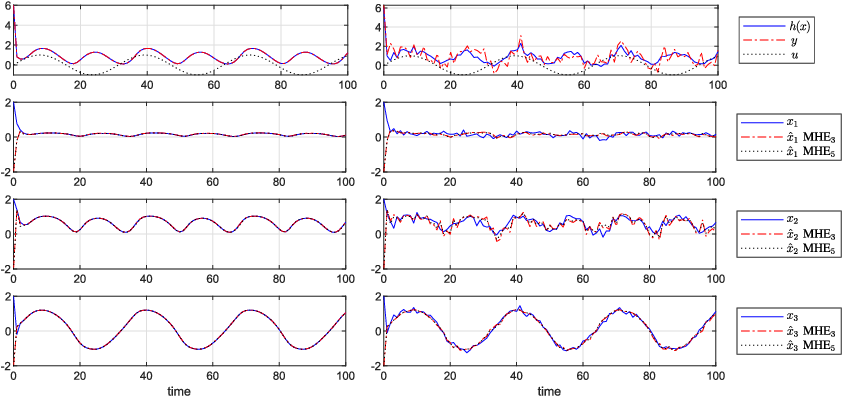}
\caption{State and estimated state variables of MHE$_3$ (with {\em fmincon} and $\varepsilon=0.01$)
and MHE$_5$ (with {\em fminimax} and $\varepsilon=0.01$)
in noise-free (on the left) and noisy (on
the right, zero-mean Gaussian noises with
dispersions equal to $0.1$ and $0.5$ for system and measurement disturbances, 
respectively) simulation runs with $x_0 = ( 2 \,,\, 2 \,,\, 2 )$,
$\hat x_0 = ( -2 \,,\, -2 \,,\, -2 )$, and $u_t=\sin (0.2 \, t)$.}
\end{figure*}

\subsection{Not Exponential i-UIOSS Example}
\label{sec:example:2}

\begin{figure*}[htb]
\label{fig:simul:example:2}
\centering
\includegraphics[width=17.5cm]{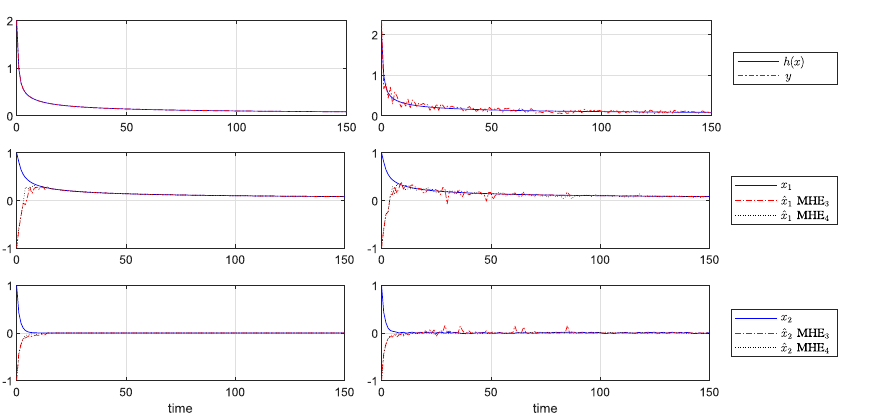}
\caption{State and estimated state variables of MHE$_3$ and MHE$_4$ with
{\em multi-start search fminimax} in noise-free (on the left) and noisy simulation
runs (on the right, with uniformly distributed random noises in 
$[-0.01,0.01]$ and $[-0.5,0.5]$ for system and measurement disturbances,
respectively) with $x_0 = ( 1 \,,\, 1 )$ and $\hat x_0 = ( -1 \,,\, -1 )$.}
\end{figure*}

The second-order system
\begin{equation}\label{eq:example:2}
\begin{array}{l}
f(x,w) = \Bigg( \dsst
\frac{x_1}{\sqrt{x_1^2+1}} + 0.1 \, x_2^2 \,,\, 
0.5 \sin ( x_2 ) + w \Bigg) \in \Re^2 \\
h(x,v) = x_1 + x_2^2 + \arctan \left( x_1 + x_2 \right) v\in \Re
\end{array}
\end{equation}
with disturbances $w, v \in \Re$ is not i-UEIOSS since the solution with 
$w \equiv 0, v \equiv 0$ and initial conditions $(\xi_0,0)$, $\xi_0 \not= 0$
is
$$
x_t = \left( \frac{\xi_0}{\sqrt{t \, \xi_0^2 +1}} \,,\, 0 
\right) \,,\, t=0,1,\ldots
$$
and thus $|x_t|$ decreases to zero slower than exponential
\cite[Example 3, p. 3419]{DucRuKell19}. 
To deal with such a system affected by the multiplicative measurement noise $v$, we need
a simple extension of \cite[Proposition 5, p. 4499]{AllanRawlACC19} for systems described by
\eqref{eq:system:general} by proving that \eqref{eq:def:i-UIOSS} in Definition \ref{def:i-UIOSS} holds
for some $\beta_x$, $\beta_y$, $\beta_v$, $\beta_w \in \cKL$ 
if there exist $V : \Bbb X \times \Bbb X \rightarrow \Re_{\ge 0}$ and $\alpha_1, \alpha_2, \alpha_3 \in \cK_\infty$,  
$\sigma_w, \sigma_y, \sigma_v \in \cK$ such that 
\begin{align}
&\alpha_1(|x-z|) \le V(x,z) \le \alpha_2(|x-z|) \label{eq:example:V:sandwich:extended} \\
&V \left( f(x,u,w_x) , f(z,u,w_z) \right) \le V(x,z) - \alpha_3(|x-z|) 
+ \sigma_w(|w_x-w_z|) + \sigma_y \left( |h(x,v_x)-h(z,v_z)| \right) \notag \\
&\qquad\qquad\qquad\qquad\qquad\qquad + \sigma_v \left( |v_x-v_z| \right) \label{eq:example:V:dissip:extended}
\end{align}
for all $x,z \in \Bbb X$, $w_x, w_z \in \Bbb W$, and $v_x, v_z \in \Bbb V$. More specifically,
there exists $\sigma \in \cK_\infty$ such that $\sigma(s) < s$ and
\begin{align}
&|x_t - z_t| \le \overbrace{\alpha_1^{-1}(\beta(\alpha_2(|x_0 - z_0|), t))}^{\beta_x(|x_0 - z_0|, t)} \oplus \Oplus_{i=0}^{t-1} 
\overbrace{\alpha_1^{-1}(\beta(\phi_w(|{w_x}_i - {w_z}_i|), t\!-\!1\!-\!i))}^{\beta_w(|{w_x}_i - {w_z}_i|, t-1-i)}
\notag \\
&\oplus \Oplus_{i=0}^{t-1} 
\overbrace{\alpha_1^{-1}(\beta(\phi_y(|h(x_i,{v_x}_i) - h(z_i,{v_z}_i)|), t\!-\!1\!-\!i))}^{\beta_y(|h(x_i,{v_x}_i)
- h(z_i,{v_z}_i)|, t-1-i)} 
\oplus \Oplus_{i=0}^{t-1} 
\overbrace{\alpha_1^{-1}(\beta(\phi_v(|{v_x}_i - {v_z}_i|), t\!-\!1\!-\!i))}^{\beta_v(|{v_x}_i
-{v_z}_i|, t-1-i)} \,,\, t=0,1,\ldots
\label{eq:i-UIOSS:example:2}
\end{align}
where $\beta(s,k) := \sigma^k(s)$,
$\phi_\ell(s) := 3 \, \alpha_2(\alpha_3^{-1}(6 \, \sigma_\ell(s))) + 3 \, \sigma_\ell(s)$ for $\ell=w, y, v$ with
$x_{t+1}=f(x_t,{w_x}_t)$,  $z_{t+1}=f(z_t,{w_z}_t)$, ${w_x}_t, {w_z}_t \in \Bbb W$, ${v_x}_t, {v_z}_t \in \Bbb V$.
The proof of \eqref{eq:i-UIOSS:example:2} is in line with the proof of \cite[Proposition 5, p. 4499]{AllanRawlACC19}
and it is omitted due to the space limitation.

To show that  $V_{c_1}\!(x,z) = \ln ( c_1 |x-z|^2 + 1)$ is an i-UIOSS
Lyapunov function for \eqref{eq:example:2} with $\sigma_{c_w}\!(s) = c_w s$, $\sigma_{c_y}\!(s) = c_y s$,
$\sigma_{c_v}\!(s) = c_v s$ and $x, z \in \Bbb X=[-2,2]\times[-2,2]$, $w_x, w_z \in \Bbb W=[-0.01,0.01]$,
and $v_x, v_z \in \Bbb V=[-0.5,0.5]$, we solved the optimization problem
\begin{align}
\max_{\tiny\begin{array}{c} \\[-0.2cm] c \ge 0, x, \! z \!\in\! \Bbb X, \\ w_x, \! w_z \!\in\! \Bbb W, v_x, \! v_z
\!\in\! \Bbb V \end{array}}  \!\!\!\!\!\!\!\! F(c,x,z,w_x,w_z,v_x,v_z) &:= \frac{1}{V_{c_1}\!(x,z)}  \Big( 
V_{c_1}\!\left(f(x,w_x),f(z,w_z)\right) -\sigma_{c_w}\!(|w_x-w_z|) \notag \\
&\quad -\sigma_{c_y}\!\left( |h(x,v_x)-h(z,v_z)| \right)  -\sigma_{c_v}\!\left( |v_x-v_z|\right) \Big) 
\label{eq:optim:V:i-UIOSS}
\end{align}
where $c := ( c_1 \,,\, c_w \,,\, c_y \,,\, c_v ) \in \Re^4_{\ge 0}$ and checked that there exists a solution
$c^*, x^*,  z^*, w_x^*, w_z^*, v_x^*, v_z^*$ such that  $c^* >0$ and $F(c^*, x^*, z^*, w_x^*, w_z^*, v_x^*, v_z^*)
\in (0,1)$. In such a case, we got $\alpha_3(s)=(1-c_0^*) \ln ( c_1^* s^2 + 1)$, where $c_0^* := F(c^*, x^*, z^*, 
w_x^*, w_z^*, v_x^*, v_z^*)$. Moreover, we chose $\alpha_1(s)=\alpha_2(s)=\ln ( c_1^* s^2 + 1)$  as
$s \mapsto \ln ( c_1^* s^2 + 1)$ is increasing over $[0,+\infty)$.
By using {\em fmincon} to solve \eqref{eq:optim:V:i-UIOSS}, we obtained $c_0^*=0.8751$, $c_1^*=6.3899$,
$c_w^* = 5.0010$, $c_y^* = 0.1822$, and $c_v^* = 4.9997$; therefore, we got $\sigma(s)=(c_0^*+1)s/2$, from which
it is straightforward to get $\beta_x(\cdot,\cdot), \beta_w(\cdot,\cdot),
\beta_y(\cdot,\cdot), \beta_v(\cdot,\cdot)$ according to the
definitions in \eqref{eq:i-UIOSS:example:2} and thus cost
\eqref{eq:cost:general}. 
The results of noise-free and noisy simulation runs are shown in Fig. 2. 

\section{Conclusion}
\label{sec:concl}

This paper presents novel insights on the robust stability
of moving horizon estimators even in case of imperfect minimization of the
cost function, which is an issue rarely addressed in the literature. 
In such a case, practical stability holds under the adoption of a suitable
stopping criterion. This allows to trade off between accuracy and computational effort.
The effectiveness of the proposed approach is illustrated by means of two case studies,
with one of the two affected by a multiplicative noise.
Future work will concern new prediction strategies in
moving-horizon estimation for Lipschitz
nonlinear systems  \cite{ArZemAlBagn23} and the robust stability of fast
moving-horizon estimators \cite{AleGagTAC17,AleGag20}.

\section*{Acknowledgements}

This work was supported by the Italian Ministry of University and Research
under Project PRIN 2022S8XSMY. The author wishes to thank Prof. Anna Rossi 
and the reviewers for the constructive comments that have allowed to improve the paper.


\end{document}